\documentclass[%
 aip,
 jmp,%
 amsmath,amssymb,
preprint%
]{revtex4-1}
\usepackage{graphicx}
\usepackage{dcolumn}
\usepackage{bm}
\usepackage{subcaption}
\usepackage{setspace}

\begin{document}

\title{Richtmyer-Meshkov Instability of an Imploding Flow with a Two-Fluid Plasma Model}

\author{Y. Li}
\author{R. Samtaney}%
 \email{ravi.samtaney@kaust.edu.sa.}
\affiliation{ 
Mechanical Engineering, King Abdullah University of Science and Technology, Saudi Arabia}%

\author{D. Bond}
\author{V. Wheatley}
\affiliation{%
Mechanical and Mining Engineering, The University of Queensland, Australia
}%

\date{\today}

\begin{abstract}
The two-fluid (ions and electrons) plasma Richtmyer-Meshkov instability of a cylindrical light/heavy density interface is numerically investigated without an initial magnetic field. Varying the Debye length scale, we examine the effects of the coupling between the electron and ion fluids. When the coupling becomes strong, the electrons are restricted to co-move with the ions and the resulting evolution is similar to the hydrodynamic neutral fluid case. The charge separation that occurs between the electrons and ions results in self-generated electromagnetic fields. We show that the Biermann battery effect dominates the generation of magnetic field when the coupling between the electrons and ions is weak. In addition to the Rayleigh-Tayler stabilization effect during flow deceleration, the interfaces are accelerated by the induced spatio-temporally varying Lorentz force. As a consequence, the perturbations develop into the Rayleigh-Taylor instability, leading to an enhancement of the perturbation amplitude compared with the hydrodynamic case. 

\end{abstract}

\maketitle

\section{\label{sec:intro}Introduction}
Inertial confinement fusion (ICF) is a promising approach for clean power generation. In ICF, a small spherical target containing fuel is expected to be compressed by an imploding wave to a hot spot of sufficiently high density and temperature for fusion ignition \citep{lindl1995}. However, hydrodynamic (HD) instabilities on the material interfaces, such as Richtmyer-Meshkov (RM) and Rayleigh-Taylor (RT) instabilities, increase the energy losses from the hot spot leading to the failure of achieving energy break-even \citep{haan2011, lindl2014}. The RM instability occurs when a perturbed interface separating fluids of different densities is accelerated by an impulse, typically, a shock wave \citep{richtmyer1960,meshkov1969}. During the shock-interface interaction, a layer of vorticity is deposited at the interface by the baroclinic source terms, leading to the growth of the initial perturbation generally divided into the following three stages: a linear stage, nonlinear stage and finally turbulent mixing \citep{zhou2017}. The instability has been extensively investigated owing to its crucial role in not only ICF but also many other fields. For instance, in astrophysics, RM instability is accounted for the lack of stratification of the products of supernova 1987A and is necessary for the stellar evolution \citep{arnett2000}; in many combustion systems, it plays a vital role in the deflagration-to-detonation transition \citep{khokhlov1999} and can be used to enhance mixing in supersonic flow \citep{yang2014}. 

Over the past decades, significant progress has been made in theory, experiments and simulations of RM instability although most of the research focuses on planar geometry \citep{zabusky1999, brouillette2002, zhai2018}. However, recently there has been considerable work on the converging RM instability motivated by its relevance to ICF. In contrast to the planar geometry, the converging shock driven RM instability occurs along with the RT effect due to the non-uniform motion or continuous acceleration of the interface \citep{lanier2003}. The RT effect is unstable and evolves into RT instability \citep{lord1883, taylor1950} when the direction of the acceleration is from heavy fluid to light one, or is stable otherwise. A reduction of the interface amplitude growth in the converging RM instability has been observed in experiments and simulations due to the RT stabilization effect \citep{lombardini2014, bakhsh2016, ding2017, lei2017}. The interface undergoes phase inversion when the RT effect is sustained for a long time \citep{luo2018}. Due to the high temperature and high energy density in ICF, the material is ionized to be in the plasma state and thus will interact with imposed and induced magnetic fields. OMEGA laser experiments have shown the potential of a strong external magnetic field to improve ICF implosion performance in terms of increasing the ion temperature \citep{chang2011, hohenberger2012}. The existence of self-generated magnetic fields has been demonstrated in ICF experiments \citep{seguin2012, igumenshchev2014}. Simulations showed that the self-generated magnetic field has an influence on the ICF implosion by affecting the electron thermal conduction \citep{srinivasan2012}. An effective fluid description for the plasma is single-fluid magnetohydrodynamics (MHD). Within this framework, in both planar and convergent geometries, theoretical and numerical investigations found that the RM instability was suppressed in the presence of an external magnetic field either parallel or normal to the interface\citep{samtaney2003, wheatley2005, wheatley2009, wheatley2014, bakhsh2016, mostert2015, mostert2017}. The mechanism of the suppression was attributed to the transport of baroclinically generated vorticity by MHD waves. The single-fluid MHD model is valid when the plasma length scales, such as Debye length and Larmor radius, are negligible compared to the characteristic length scale of the flow. However, magnetized implosion experiments have demonstrated that the Larmor radius of alpha particles may be larger than the hot spot size \citep{hohenberger2012}, suggesting that single-fluid MHD may not be sufficient to model the physics under these circumstances. Besides, unless the Biermann battery effect is included \citep{srinivasan2012}, MHD fails to capture the phenomenon of self-generated magnetic fields. To consider the effect of plasma length scales, the two-fluid plasma model is employed in which ions and electrons are treated as two separate fluids and are coupled to the full Maxwell equations. The charge separation and self-generated electromagnetic field may then be investigated through the two-fluid plasma model. It is noted that the electron inertia and light speed are finite in two-fluid plasma equations unlike the Hall MHD model. Bond {\it{et al.}} have investigated the two-fluid plasma single-mode RM instability in planar geometry \citep{bond2017} and showed that the self-generated electromagnetic field increased the interfacial growth rates and caused high wavenumber instabilities, which may be more detrimental to ICF than those predicted by single-fluid models. As an extension of this work, we study the converging RM instability with two-fluid plasma model in this paper. The imploding shocks that impact the perturbed density interfaces are generated via a Riemann problem in this paper. By varying the plasma length parameter (Debye length), we discuss the coupling effect on the flow. We show that the perturbation amplitude growth increases compared with HD case, due to the electromagnetic driven RT instability. As the coupling becomes strong, the extent of the perturbation growth enhancement decreases. We show that the Biermann battery effect is one main contribution to the magnetic field generation for the whole time when coupling is weak. However, it dominates only during the period of ion shock-interface interaction in the strong coupling situation.

The remainder of the paper is organized as follows: The ideal two-fluid plasma equations are introduced in Section \ref{sec:ITFP}, followed with the initial setup of the problem and a brief description of the numerical method in Section \ref{sec:setup}. In Section \ref{sec:results}, simulation results are presented: herein we focus on the evolution of flow field and electromagnetic field and quantify the growth of perturbation and compared with that of the HD case. Moreover, the Biermann battery effect on the self-generated magnetic field is discussed. Conclusions are presented in Section \ref{sec:conclusions}.
\section{Ideal two-fluid plasma model}\label{sec:ITFP}
An ideal two-fluid plasma model is applied in this study. In this model, the ions and electrons are described by Euler equations with Lorentz forces as the source term,
\begin{eqnarray}
&&\frac{\partial \rho_\alpha}{\partial t}+\nabla \cdot \left(\rho_\alpha \bm{u}_\alpha \right) = 0, \\
&&\frac{\partial \rho_\alpha \bm{u}_\alpha}{\partial t}+\nabla \cdot \left(\rho_\alpha \bm{u}_\alpha \bm{u}_\alpha +p_\alpha \bm{I} \right) = n_\alpha q_\alpha \left(\bm{E}+\bm{u}_\alpha \times \bm{B} \right),\\
&&\frac{\partial \cal{E}_\alpha}{\partial t}+\nabla \cdot \left(\left({\cal{E}_\alpha} + p_\alpha \right)\bm{u}_\alpha\right) = n_\alpha q_\alpha \bm{E} \cdot \bm{u}_\alpha,
\end{eqnarray}
where,
\begin{equation}
\rho_\alpha=n_\alpha m_\alpha, \quad p_\alpha=n_\alpha k_B T_\alpha, \quad {\cal{E}_\alpha}=\frac{p_\alpha}{\gamma_\alpha-1}+\frac{\rho_\alpha |\bm{u}_\alpha|^2}{2}.
\end{equation}
The subscript $\alpha$ denotes the species with `$\alpha=i (e)$' for ions (electrons). $\rho$, $n$, $m$, $\bm{u}=\left(u, v, w \right)^\text{T}$, $p$, $\cal{E}$, $q$ and $T$ are the density, number density, particle mass, velocity, pressure, energy, particle charge and temperature, respectively. $\gamma$ is specific heat ratio and $k_B$ is the Boltzmann constant. In the model, the collisions between particles are not considered, thus ions and electrons interact solely via electromagnetic forcing terms. The evolution of magnetic field $\bm{B}$ and electric field  $\bm{E}$ is governed by the Maxwell equations. Because of discretization errors, the errors in the electromagnetic divergence constraints increase with time. To numerically satisfy the divergence constraints, two Lagrangian multipliers $\psi_B$ and $\psi_E$ are introduced \citep{Munz2000},
\begin{eqnarray}
&&\frac{\partial \bm{B}}{\partial t}+\nabla \times \bm{E} +\Gamma_B \nabla \psi_B= \bm{0}, \\
&&\frac{\partial \bm{E}}{\partial t}- c^2 \nabla \times \bm{B} +c^2\Gamma_E \nabla \psi_E= -\frac{1}{\epsilon_0} \sum_{\alpha}n_\alpha q_\alpha \bm{u}_\alpha, \\
&&\frac{\partial \psi_E}{\partial t}+\Gamma_E \nabla \cdot \bm{E} = \frac{\Gamma_E}{\epsilon_0}\sum_{\alpha}n_\alpha q_\alpha, \\
&&\frac{\partial \psi_B}{\partial t}+c^2 \Gamma_B\nabla \cdot \bm{B} = 0,
\end{eqnarray}
where $c=1/\sqrt{\mu_0 \epsilon_0}$ is the light speed, $\mu_0$ is permeability of free space and $\epsilon_0$ is vacuum permittivity. As a result, the introduced correction potentials $\psi_B$ and $\psi_E$ transfer the divergence constraint errors out of the domain with the speed $\Gamma_B c$ and $\Gamma_E c$, respectively. Here, $\Gamma_B $ and $\Gamma_E $ are chosen to be 1 throughout this study. 

By specifying the reference variables (with subscript $0$), the dimensionless variables are defined as:
\begin{eqnarray}
&&\hat{\bm{x}}=\frac{\bm{x}}{L_0},~ \hat{t}=\frac{t}{L_0/u_0},~ \hat{\rho}_\alpha=\frac{\rho_\alpha}{n_0m_0},~ \hat{m}_\alpha=\frac{m_\alpha}{m_0},~ \hat{\bm{u}}_\alpha=\frac{\bm{u}_\alpha}{u_0},~ \hat{q}_\alpha=\frac{q_\alpha}{q_0},~ \hat{p}=\frac{p_\alpha}{n_0 m_0 u_0^2},~\nonumber \\
&&\hat{\bm{B}}=\frac{\bm{B}}{B_0},~ \hat{\bm{E}}=\frac{\bm{E}}{c B_0},~ \hat{\psi}_E=\frac{\psi_E}{B_0},~ \hat{\psi}_B=\frac{\psi_B}{c B_0},~ \hat{c}=\frac{c}{u_0},
\end{eqnarray}
where the reference magnetic field $B_0=\sqrt{\mu_0 n_0 m_0 u_0^2}$. Therefore, the dimensionless ideal two-fluid plasma equations with the above notation may be written as follows, with the carets omitted for simplicity,
\begin{eqnarray}
\label{eq:density}
&&\frac{\partial \rho_\alpha}{\partial t}+\nabla \cdot \left(\rho_\alpha \bm{u}_\alpha \right) = 0, \\
\label{eq:momentum}
&&\frac{\partial \rho_\alpha \bm{u}_\alpha}{\partial t}+\nabla \cdot \left(\rho_\alpha \bm{u}_\alpha \bm{u}_\alpha +p_\alpha \bm{I} \right)= \frac{n_\alpha q_\alpha}{d_{D,0} c} \left(c\bm{E}+\bm{u}_\alpha \times \bm{B} \right),\\
\label{eq:energy}
&&\frac{\partial \cal{E}_\alpha}{\partial t}+\nabla \cdot \left(\left({\cal{E}_\alpha} + p_\alpha \right)\bm{u}_\alpha\right) =  \frac{n_\alpha q_\alpha }{d_{D,0}} \bm{E} \cdot \bm{u}_\alpha,\\
\label{eq:magnetic}
&&\frac{\partial \bm{B}}{\partial t}+c\nabla \times \bm{E} +c\Gamma_B \nabla \psi_B= \bm{0}, \\
\label{eq:electric}
&&\frac{\partial \bm{E}}{\partial t}- c \nabla \times \bm{B} +c\Gamma_E \nabla \psi_E= -\frac{1}{d_{D,0}} \sum_{\alpha}n_\alpha q_\alpha \bm{u}_\alpha, \\
\label{eq:psi_E}
&&\frac{\partial \psi_E}{\partial t}+c\Gamma_E \nabla \cdot \bm{E} = \frac{c \Gamma_E}{d_{D,0}}\sum_{\alpha}n_\alpha q_\alpha, \\
\label{eq:psi_B}
&&\frac{\partial \psi_B}{\partial t}+c \Gamma_B\nabla \cdot \bm{B} = 0.
\end{eqnarray}
Here $d_{D,0}=\sqrt{\frac{\epsilon_0 m_0 u_0^2}{n_0 q_0^2 L_0^2}}$ is the reference Debye length. 
The scale of $d_{D,0}$ reveals the coupling between ions and electrons in some extents, the larger $d_{D,0}$ is , the smaller the coupling between the two species. 
Decreasing the Debye length implies stronger  coupling effect and the limiting behavior of $d_{D,0}\rightarrow 0$ is that the two species evolve together as a ``single" fluid.

\section{Initial setup}\label{sec:setup}
\begin{figure}
\includegraphics[width=\linewidth]{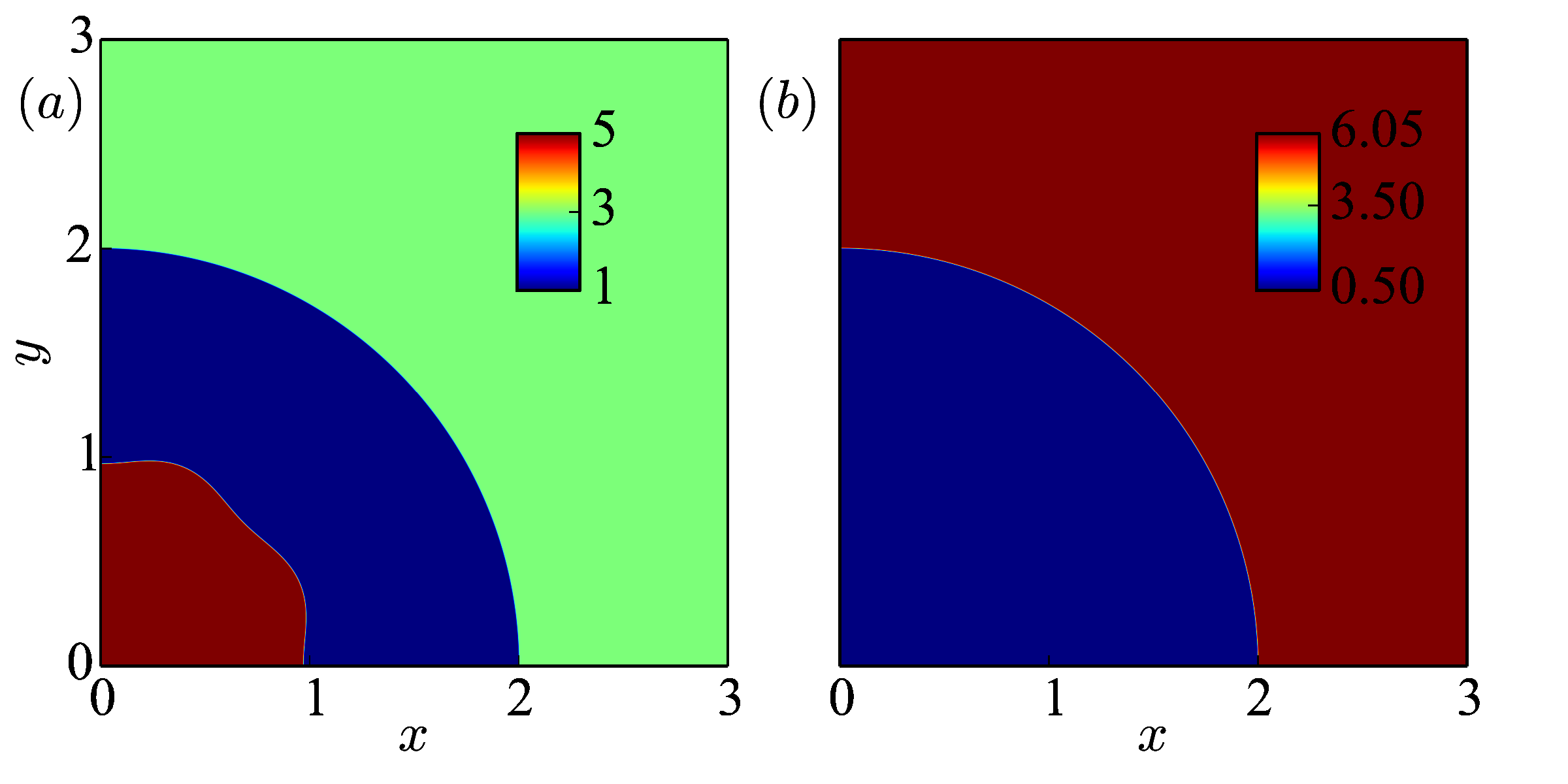}
\caption{\label{fig:setup} Initial conditions of ions and electrons: (a) number density, (b) pressure.}
\end{figure}

Fig.~\ref{fig:setup} shows the initial conditions for the converging RMI. The density interface radial location, perturbed with a single-mode of azimuthal wavenumber $k$, is 
\begin{equation}
\zeta_0(\theta)=r_0-\eta_0 \cos(k\theta),
\end{equation}
where $r_0$ is the mean radius of the density interface and $\eta_0$ is the amplitude of the perturbations. In this study, $r_0$, $k$, and $\eta_0$ are set to be $1$, $8$ and $4\%$ of the perturbation wavelength, respectively. A Riemann interface centered at the origin with $r_I=2$ is applied to generate converging shocks that interacts with the perturbed interface. As seen from Fig.~\ref{fig:setup}, the whole domain is divided into three sections. From inside to outside, the initial (non-dimensional) number density $n$ and pressure $p$ of each species in each sections are,
\begin{equation}
\left(
\begin{array}{c}
n\\
p
\end{array}\right)_{i,e}=\left(
\begin{array}{c}
5\\
0.5
\end{array}\right), \quad
\left(
\begin{array}{c}
1\\
0.5
\end{array}\right), \quad
\left(
\begin{array}{c}
3\\
6.05
\end{array}\right),
\end{equation}
while the initial velocities are set to be $0$. In each section,  charge neutrality, thermal equilibrium and mechanical equilibrium are satisfied initially. A hyperbolic tangent density profile is defined across the density interfaces, 
\begin{equation}
n(r,\theta)=\frac{1}{2}\left(n_{L}+n_{R}+\left(n_R-n_L\right)\tanh\left(\mu\left(r-\zeta_0(\theta)\right)\right)\right),
\end{equation}
where $\mu=500$. The non-dimensional electron charge $q_e$, ion charge $q_i$, electron mass $m_e$  and ion mass $m_i$ are set to be $-1$, $1$, $0.01$ and $1$, respectively. Here we use the same mass ratio $m_i/m_e=100$ as in Bond {\it{et al.}} \citep{bond2017} instead of the physical value $1836$ to reduce numerical stiffness. Based on the limiting values of hotspot temperature and number density in ICF implosion ($T_0=5 \times 10^3$ eV with $n_0=10^{31}$) \citep{srinivasan2012}, the value of non-dimensional light speed $c$ was chosen to be $50$ to alleviate the computational cost. In this work, $\gamma_\alpha$ is $5/3$ for each species. It is noted that no initial electromagnetic field is applied. Under the limits of the reference light speed $c \rightarrow \infty$ and $d_{D,0}c \rightarrow 0$ \citep{shen2018}, the ideal two-fluid plasma model equations approach the ideal MHD model, while the ideal MHD model is identical to HD if no initial magnetic field is present. Therefore, the two-fluid plasma results are expected to approach the HD case  under the same limits. In our work, it turns out the initial density and pressure in each section of the corresponding HD case are $\sum_\alpha \rho_\alpha$ and $\sum_\alpha p_\alpha$, respectively. The two-fluid plasma flow is investigated under different reference Debye lengths. We also compare the results with those of the corresponding HD case.  

A second-order finite-volume numerical method is used to solve the equations, with HLLC solver for the fluid fluxes and HLLE scheme for the electromagnetic fluxes. The source terms are treated locally with an implicit method \citep{abgrall2014}. For time discretization, the strong stability preserving Runge-Kutta scheme is applied \citep{gottlieb2001}. Verification of this solver refers to 	the paper by Bond {\it{et al.}} \citep{bond2017}. All the simulations in the work are computed with the resolution of $2048$ cells per unit length. Reflected conditions are applied at inner boundaries ($x=0$, $y=0$) while outflow conditions are used on the outer edges ($x=3$, $y=3$). A volume-of-fluid approach is used to track the density interface, where the tracer $\phi_\alpha \in\left[0,1\right]$ is for each species.

\section{\label{sec:results}Results and discussions}
Presently, we investigate the converging RM instability with the two-fluid plasma model. The converging wave structures generated from the initial Riemann interface in the ion and electron fluids are shown using a space-time diagram in \ref{sec:baseflow}. The interactions between these waves and perturbed interfaces are considered in \ref{sec:flowevol}. We first study the case with a relatively large reference Debye length $d_{D,0}=0.1$, where the two-fluid effects exert a significant influence on the flow evolution. Following this, we investigate the evolution of the self-generated electromagnetic field in \ref{sec:selfemag}. Then we study the case of reference Debye length $d_{D,0}=0.01$ which supposed to be closer to the single-fluid MHD limit in \ref{sec:smallDebye}. Finally, the evolution of perturbations is discussed in \ref{sec:perturbations}.

\subsection{\label{sec:baseflow}Converging wave structures}
\begin{figure}
\includegraphics[width=1\linewidth]{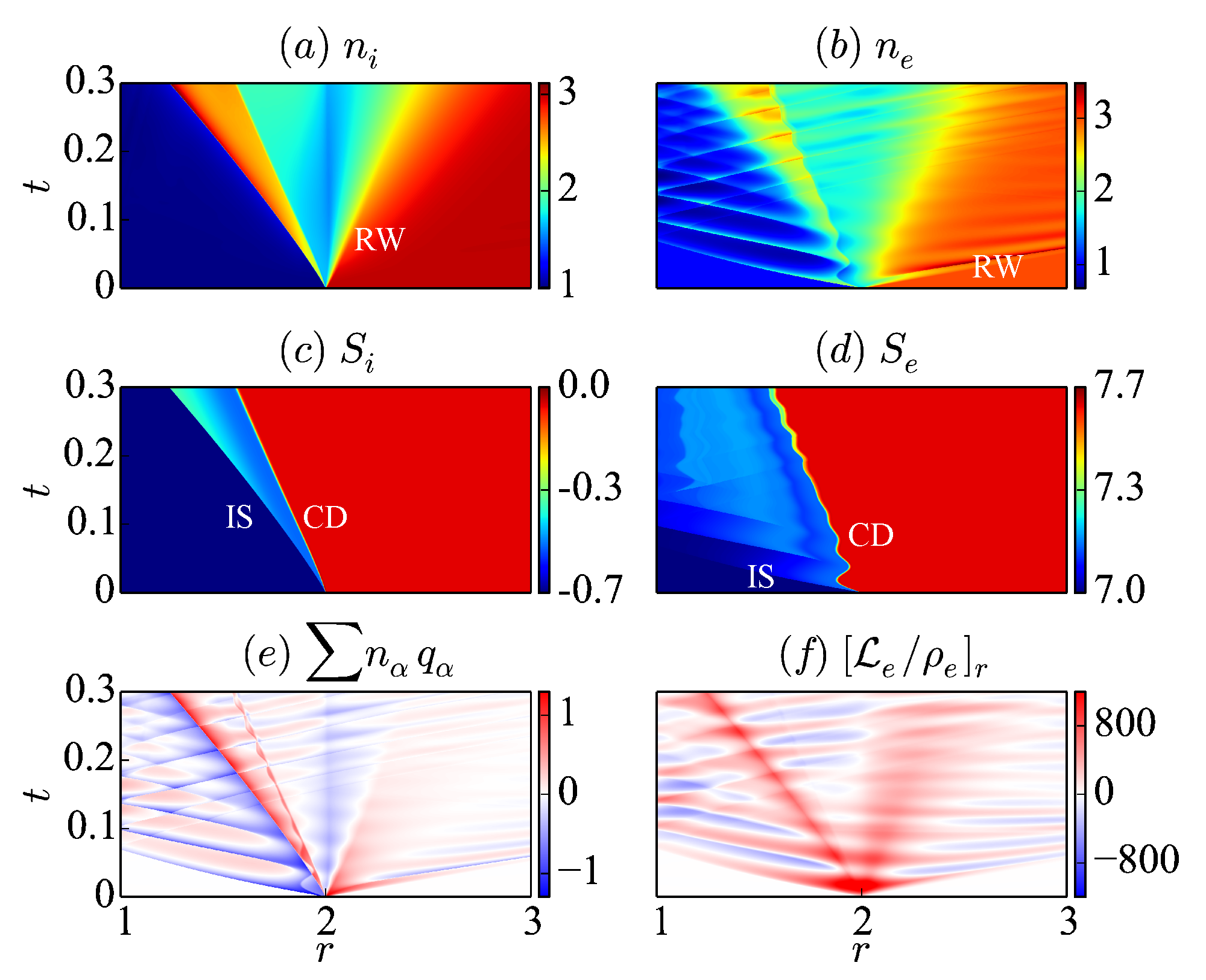}
\caption{\label{fig:xt_dd0_1} Space-time diagrams along $r$ direction with $d_{D,0}=0.1$. (a) ion number density, (b) electron number density, (c) ion entropy $S_i=\ln{(p_i/\rho_i^{\gamma_i})}$, (d) electron entropy $S_i=\ln{(p_e/\rho_e^{\gamma_e})}$, (e) charge density, (f) Lorentz acceleration of electrons in $r$ direction. `IS': incident shock, `CD': contact discontinuity, `RW': rarefaction wave.}
\end{figure}
We consider the case with $d_{D,0}=0.1$. Due to the relatively large reference Debye length, the coupling effect between ions and electrons is relatively weak and the wave structures differ significantly between the two species. In order to study the interaction between the density interfaces and converging waves, we first investigate the imploding wave structures generated from the Riemann interface in ions and electrons through space-time diagrams along the radial direction in the range of $1<r<3$ for $t\le 0.3$, as shown in Fig. \ref{fig:xt_dd0_1}. The ions depict a relatively simple wave structure: a radially converging shock, a contact discontinuity and a rarefaction wave, generated from the Riemann interface, see Fig. \ref{fig:xt_dd0_1}(a,c). In contrast, a much more complicated wave structure occurs in electrons, with multiple converging shocks and an oscillating contact discontinuity, as shown in Fig. \ref{fig:xt_dd0_1}(b,d). Owing to an order of magnitude higher sound speed in electrons, the converging shocks in electrons travel much faster than the ion shock. As a result, at early time $t\approx0.08$, the ion converging wave has not travelled far while the first electron converging wave has interacted with the electron interface. By $t=0.3$, multiple electron waves have been reflected from the electron interface while the ion shock has not yet impinged on the ion interface. 
The relative motion between contact discontinuities in the electron and ion fluids leads to charge separation with a corresponding Lorentz force. Fig. \ref{fig:xt_dd0_1}(c,d) show that the induced Lorentz force has little effect on ions while it exerts a significant influence on the electrons due to the much smaller electron mass. A clear trace of the ion waves in the electron number density plot can be seen in Fig. \ref{fig:xt_dd0_1}(b). Since the induced electromagnetic force aims to reduce the charge separation, it impedes the movement of the electrons on the contact discontinuity and pulls the electron contact interface back towards the ion contact interface. As a result, the travel direction of the electron contact discontinuity is reversed by the initial strong positive acceleration which is manifested as the first oscillation, as depicted in \ref{fig:xt_dd0_1}(d). Subsequently, the reversed electron contact interface overshoots to the right of the ion contact discontinuity, resulting in the positive radial Lorentz force on it. As a consequence, it is pulled back and forms the second period oscillation. Fig. \ref{fig:xt_dd0_1}(b,d) shows that  this process continues while the oscillation amplitude decreases until the reflected waves from the electron interface interact with the contact discontinuity and increase the oscillation amplitude again.  Each inward oscillation of the electrons drives a converging shock into the electron fluid, as indicated in Fig. \ref{fig:xt_dd0_1}(b). Due to the compression effect of the converging electron shocks, the downstream number density of electrons increases while the ions are mostly unperturbed, thus leading to a negative charge density behind the electron shocks. In addition, since the downstream electrons are compressed towards to the electron shocks such that the remain charge of downstream area becomes positive to maintain overall charge neutrality, therefore we observe (see Fig. \ref{fig:xt_dd0_1}(e)) alternate negative and positive charge bands behind the electron shocks. 
\subsection{\label{sec:flowevol}Flow evolution}
\begin{figure}
\includegraphics[width=0.7\linewidth]{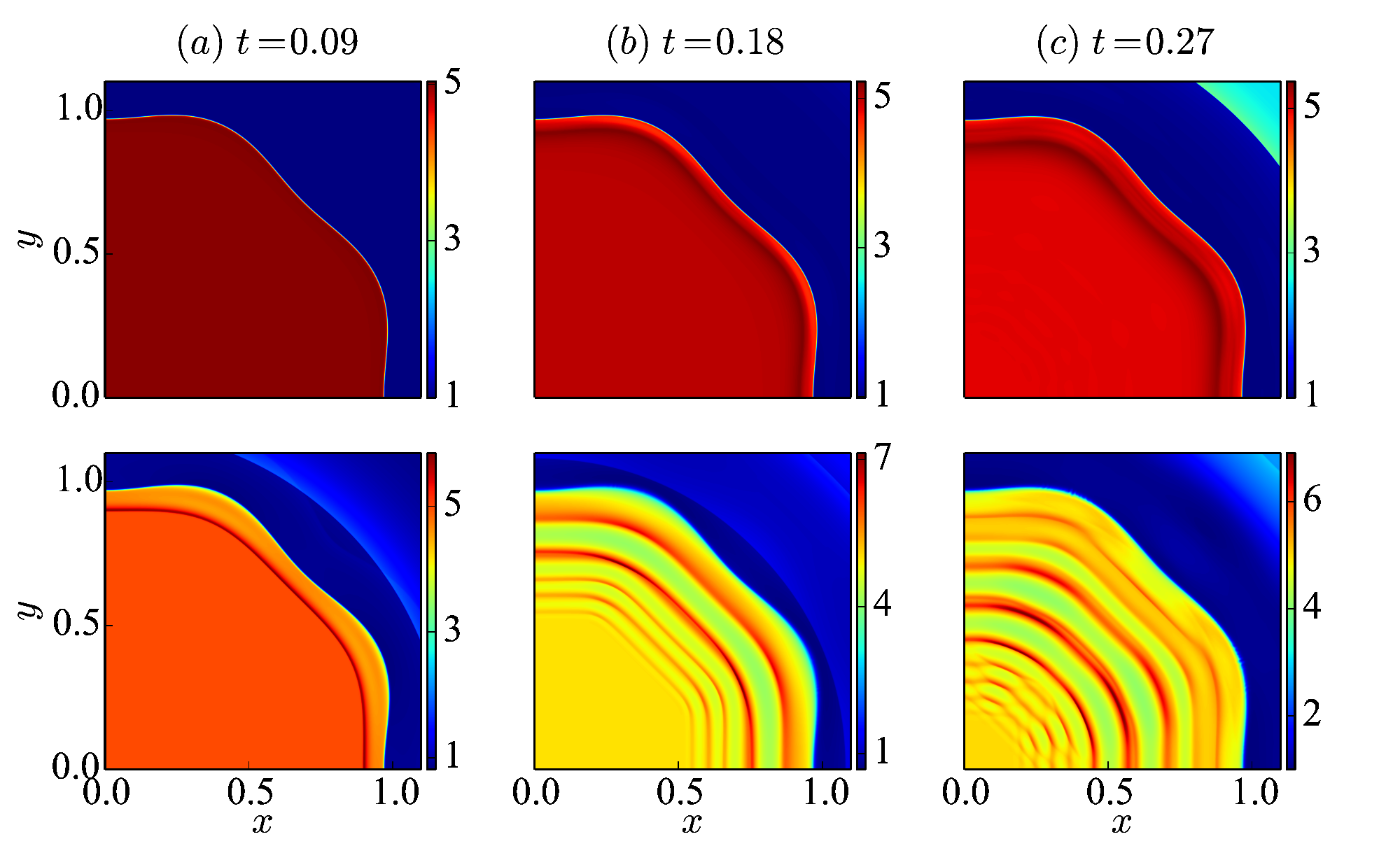}\\
\includegraphics[width=0.7\linewidth]{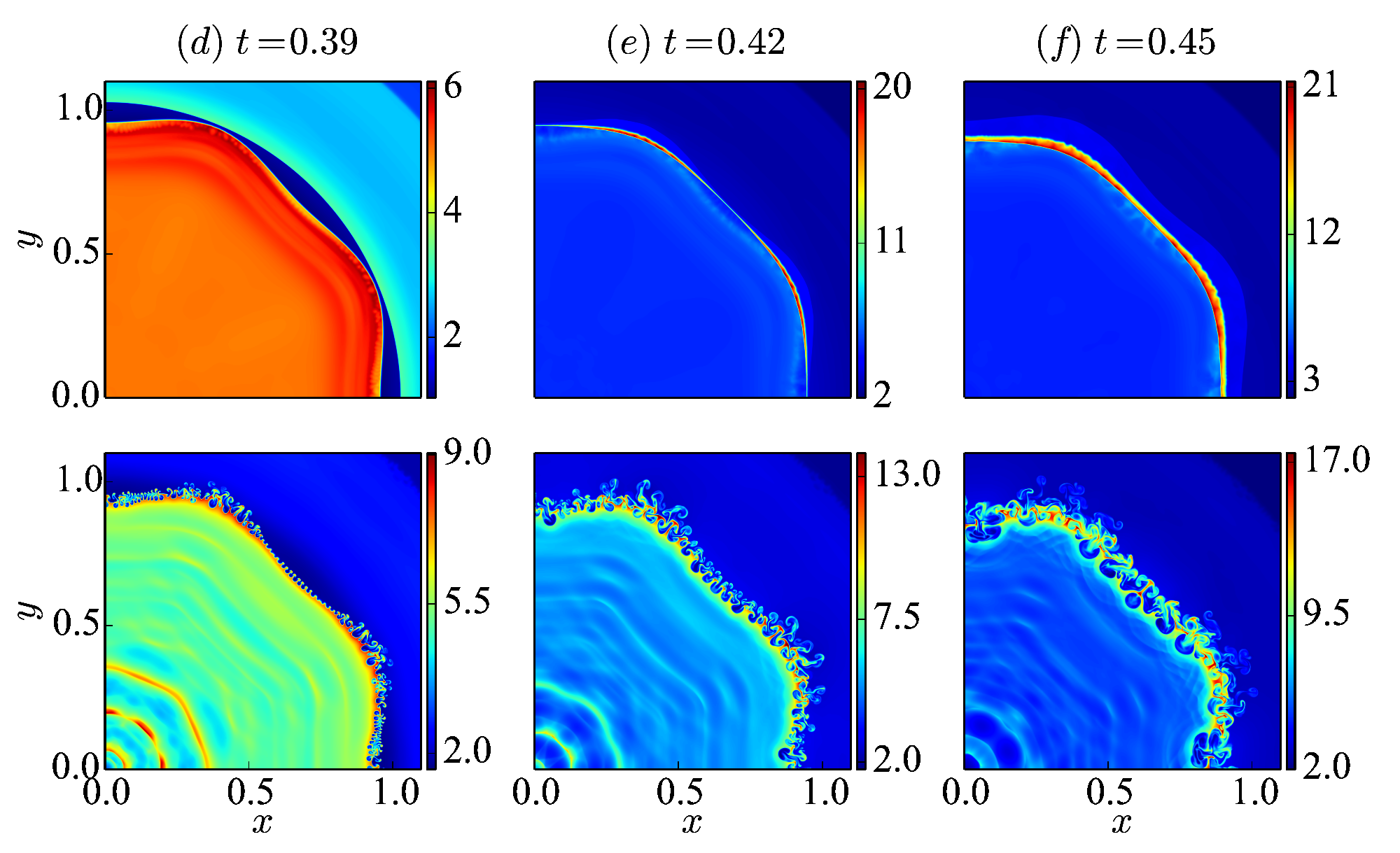}\\
\includegraphics[width=0.7\linewidth]{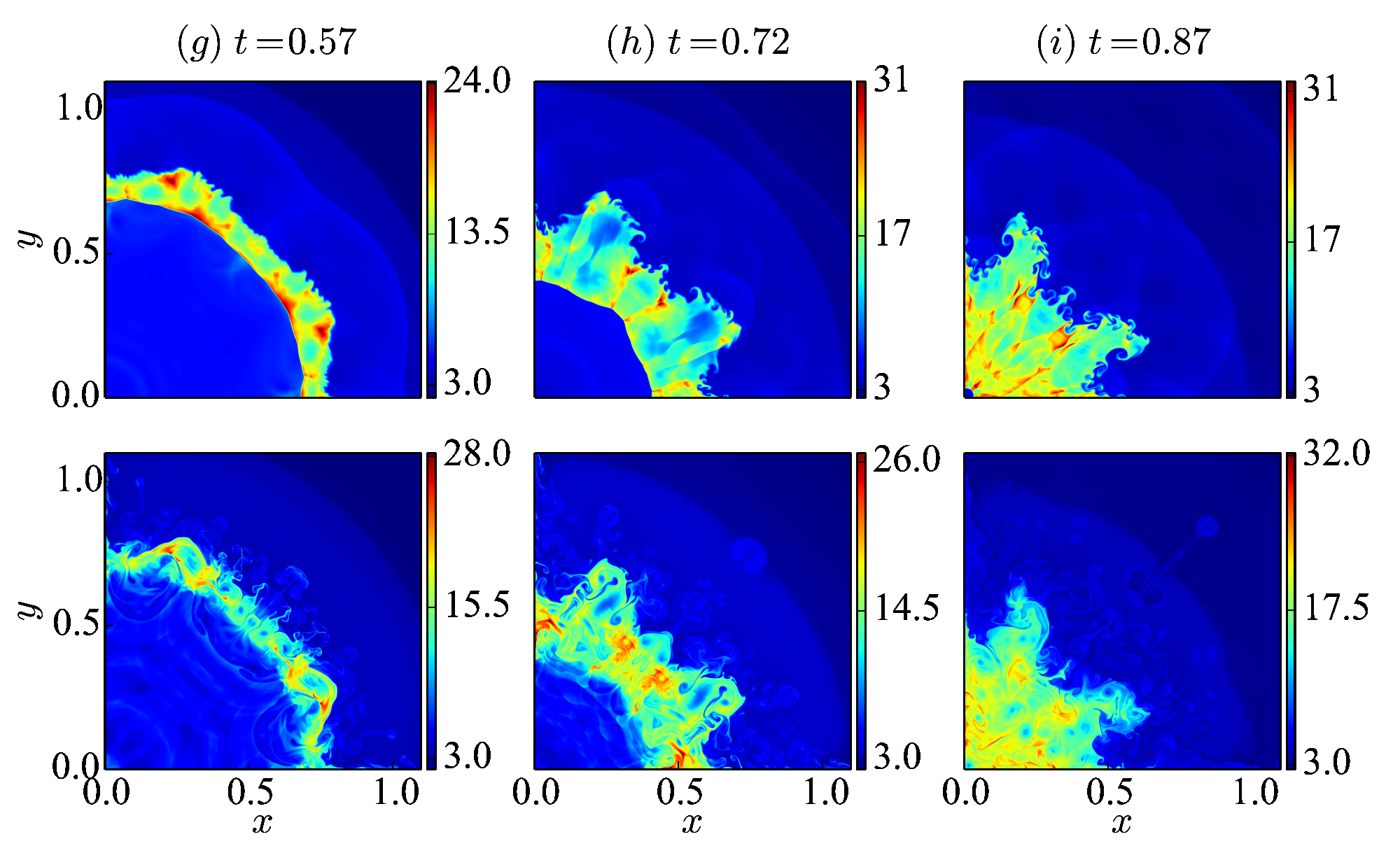}
\caption{\label{fig:numden_dd0_1} Number density of ions (top) and electrons (bottom) at various time with $d_{D,0}=0.1$.}
\end{figure}
\begin{figure}
\includegraphics[width=1\linewidth]{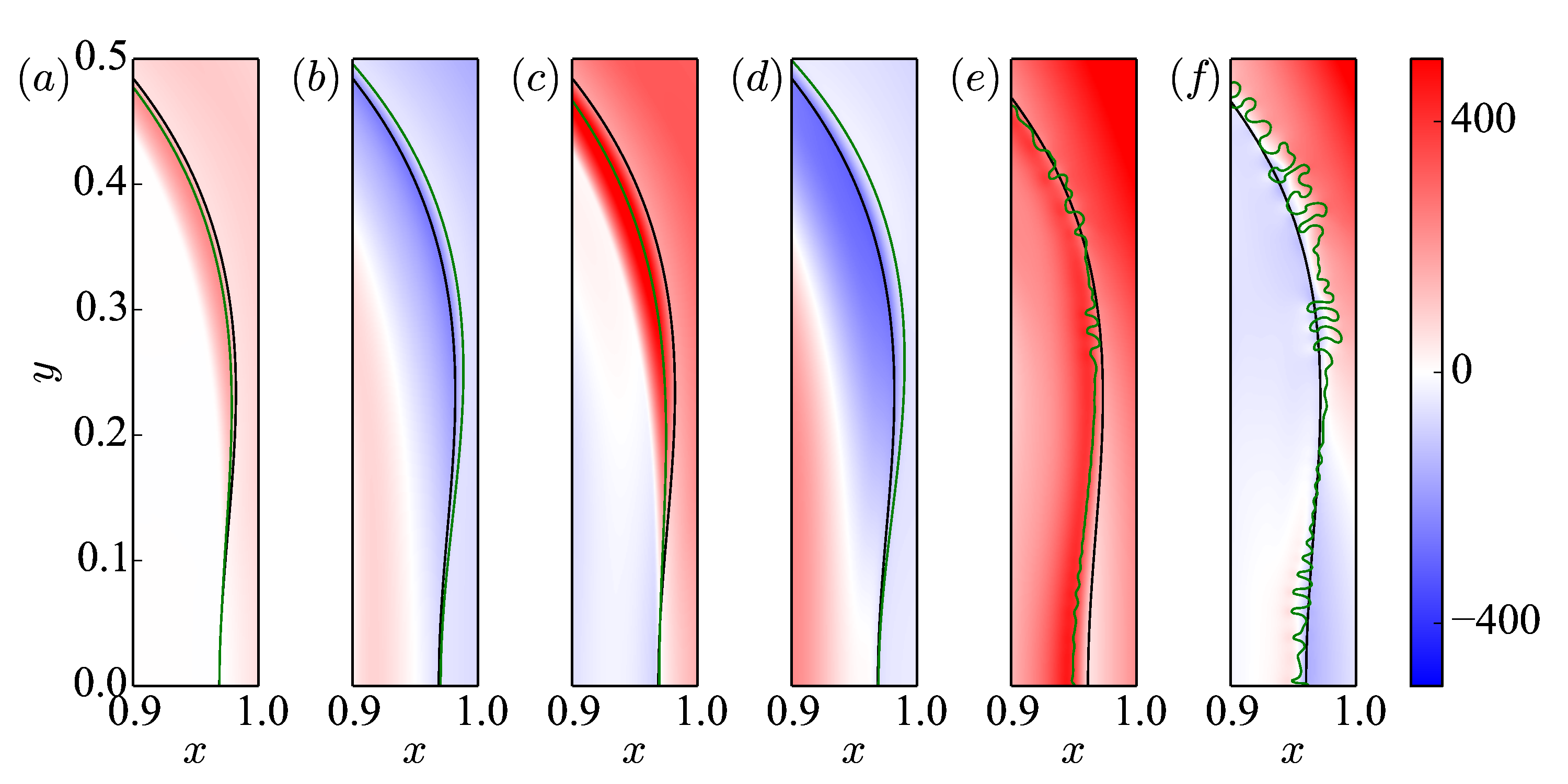}
\caption{\label{fig:DI_dd0_1}  Radial electron Lorentz acceleration $[\mathcal{L}_e/\rho_e]_r$ for the case with $d_{D,0}=0.1$ at various time overlaid with density interface of ions (black) and electrons (green). (a) $t=0.075$, (b) $t=0.09$, (c) $t=0.105$, (d) $t=0.12$, (e) $t=0.345$, (f) $t=0.36$}
\end{figure}
\begin{figure}
\includegraphics[width=1\linewidth]{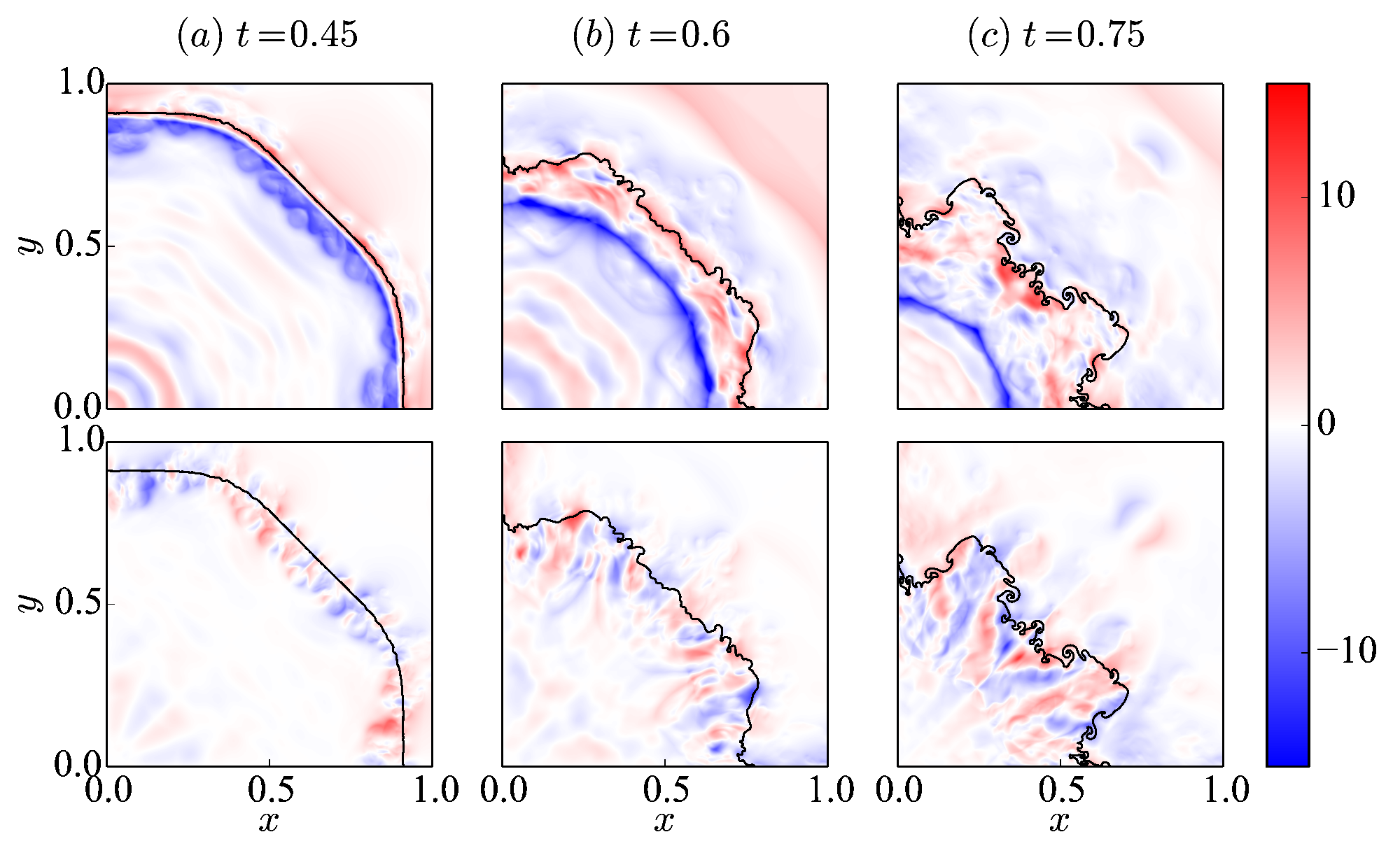}
\caption{\label{fig:Li_rhoi_dd0_1}  Ion Lorentz acceleration $[\mathcal{L}_i/\rho_i]$ along radial (top) and  azimuthal (bottom) direction for the case with $d_{D,0}=0.1$ at various time overlaid with ion interface (black line).}
\end{figure}
Next we discuss the flow evolution after the interaction between imploding waves and perturbed interfaces. Fig. \ref{fig:numden_dd0_1} shows the evolution of number density of the two fluids. As shown in the plots, multiple shock-interface interactions have occurred in electrons before the ion interface is impacted by the ion shock at $t \approx 0.39$. The perturbed transmitted electron shocks form a wave packet that converge to the origin while the strength of the converging waves are weakened by the induced Lorentz force, as shown in Fig. \ref{fig:numden_dd0_1}(b). During the convergence, the curvatures of the first converging electron shock increases eventually leading to triple points and formation of shock-shocks. The reflected shocks from such shock-shock interactions influence the ensuing electron converging shocks in the wave packets, as depicted in Fig. \ref{fig:numden_dd0_1}(c). Since the coupling effect between the two fluids is weak with the relatively large Debye length, the ions are almost unaffected at early time ($t=0.09$ Fig. \ref{fig:numden_dd0_1}(a)). However, the electron dynamics exerts a long-term influence on the ions as evident from the compression zone near the ion interface at $t=0.27$. Fig. \ref{fig:DI_dd0_1} shows the evolution of density interfaces before the ion shock arrives. At $t=0.075$, the electron interface has been compressed by the first electron shock, resulting in the outward Lorentz force on electrons near the interface. The induced Lorentz force accelerates the electrons towards the ion interface, as a result of which the electron interface moves back and overshoots to the right of the ion interface by $t=0.09$ with a negative electron Lorentz acceleration generated in the meantime. Then, under the compression effect due to the second imploding electron shock and the inward drag effect of Lorentz force, the electron interface reverses and moves radially inward to the ion interface at $t=0.105$ and finally travels radially outward from the ion interface at $t=0.12$ due to the Lorentz force. This process continues under the combined effects of imploding electron shocks and induced electromagnetic force, with the  consequence that the electron interface oscillates about the ion interface. The ions are influenced under an almost equal and opposite electromagnetic force, but it is noted that the oscillations of ion interface are negligible due to the long response time and much larger inertia of the ions. Later, the spatio-temporal variation of the electron Lorentz accelerations drive the secondary RT instability on electron interface, as shown in Fig.\ref{fig:DI_dd0_1}(e,f), where we note the development of the fine-scale finger-shape perturbations on the electron interface. At $t=0.39$, the ion shock with strength of Mach $\approx 2.9$ arrives at the ion interface and traverses it by $t=0.45$. During this time, the compressed ion interface remains relatively smooth while the secondary instabilities on electron interface rapidly grow and merge that make the electron interface chaotic, see Fig. \ref{fig:numden_dd0_1}(d,e,f). After passing over the interface, the ion shock is perturbed with the same primary wavenumber as the ion interface. However, it is also perturbed by the sufficiently strong electromagnetic force acting on it, and hence we observe secondary oscillations generated on the shock at $t=0.57$. As the ion shock converges, the curvatures of these perturbations on ion shock steepens and form triple points. Subsequently, these shock-shock interactions generate complex shaped high density regions behind the converging shock in ions, as indicated in Fig. \ref{fig:numden_dd0_1}(i). It shows that the RM instability develops after the shock-interface interaction. The primary modes of perturbations grow and form two spikes by $t=0.87$. In addition to the RM instability, the RT instability of ion interface grows over a long duration time of the electromagnetic acceleration. Fig. \ref{fig:Li_rhoi_dd0_1} shows the ion Lorentz acceleration after the interface is impacted by the ion shock. We observe that the incident ion shock creates a long-lived positive Lorentz force acting on the ion interface in radial direction which is the RT unstable direction (from heavy to light) while a similar effect is found in the azimuthal direction. Due to the effect of long-term, space-varying accelerations in the RT unstable directions, the high-wavenumber secondary instabilities of ion interface are manifested. For instance, a small-scale spike has developed along the line $y=x$ by $t=0.75$, as shown in Fig. \ref{fig:Li_rhoi_dd0_1}(c). The combination of RM and RT instabilities cause a larger perturbation growth of the ion interface in two-fluid plasma than the single-fluid RM instability case, discussed later in \ref{sec:perturbations}. Although the morphology of the flow field is significantly different between the two fluids at the early stage, the general structures of flow field become similar at late times, since the induced Lorentz force tends to reduce the charge separation. However, more fine-scale structures are exhibited in electrons due to the light particle mass. 

\subsection{\label{sec:selfemag}Field evolution}
\begin{figure}
\includegraphics[width=\linewidth]{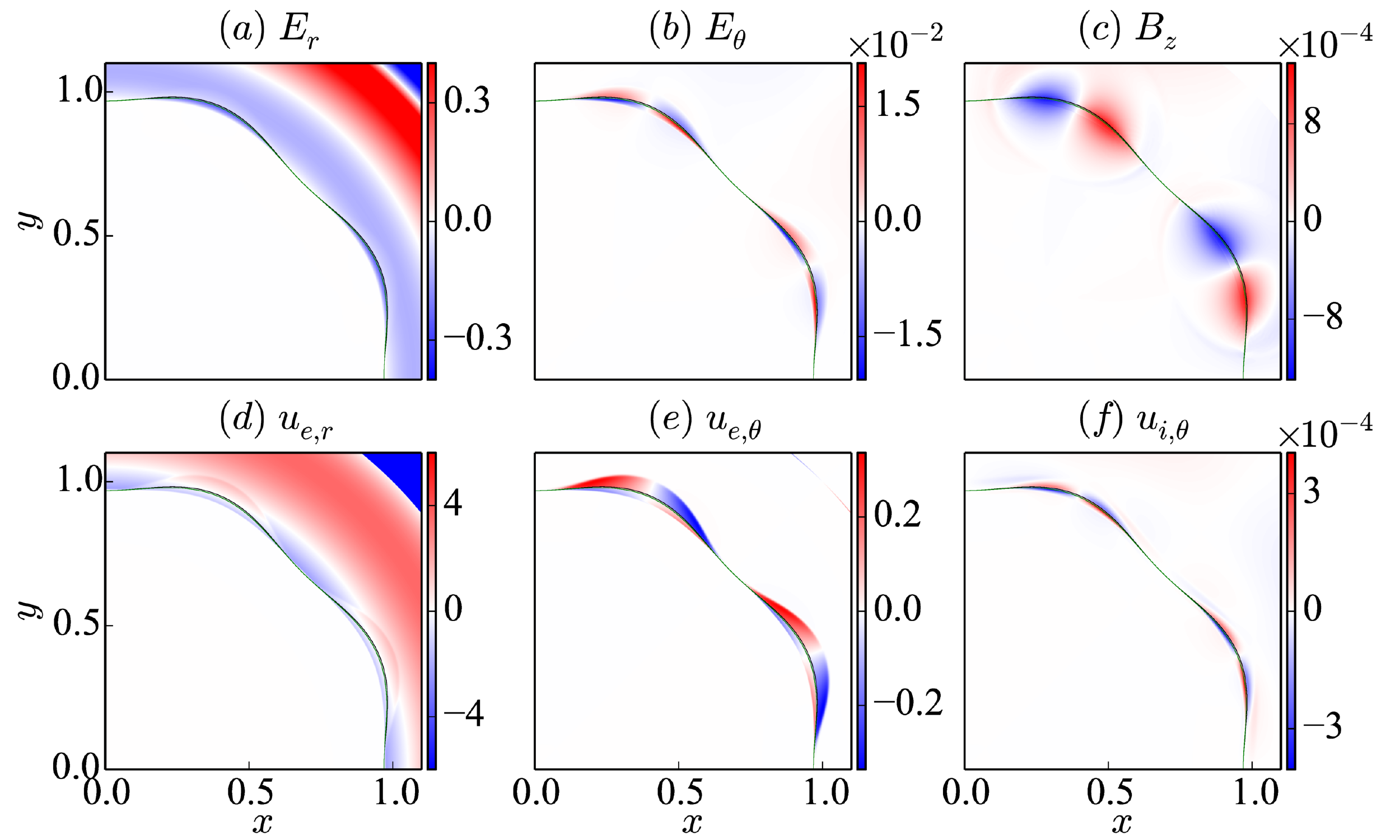}
\caption{\label{fig:fieldGen_dd0_1} Generated field and velocities overlaid with ion DI (black) and electron DI (green) at $t=0.075$.}
\end{figure}
\begin{figure}
\includegraphics[width=0.7\linewidth]{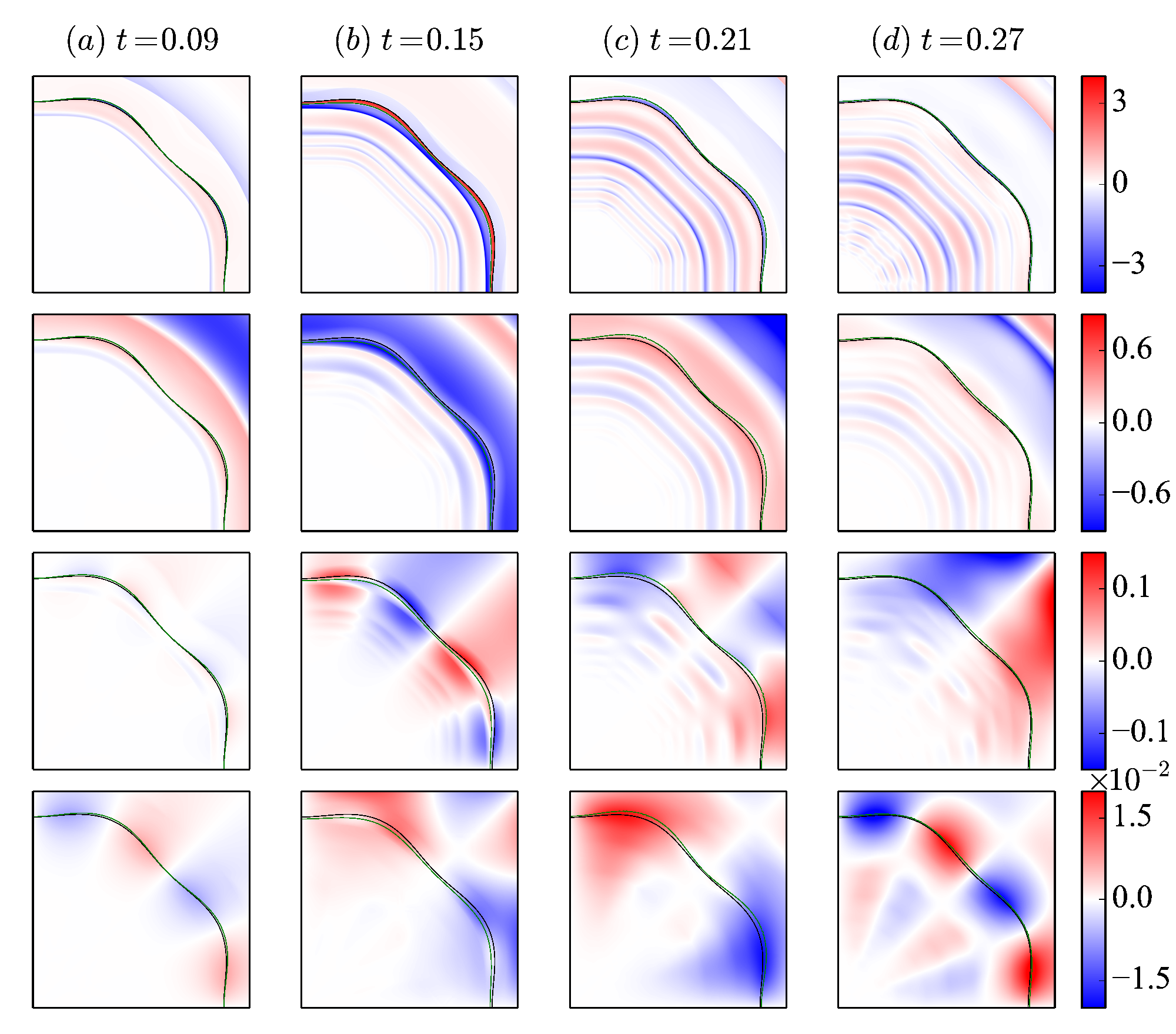}\\
\includegraphics[width=0.7\linewidth]{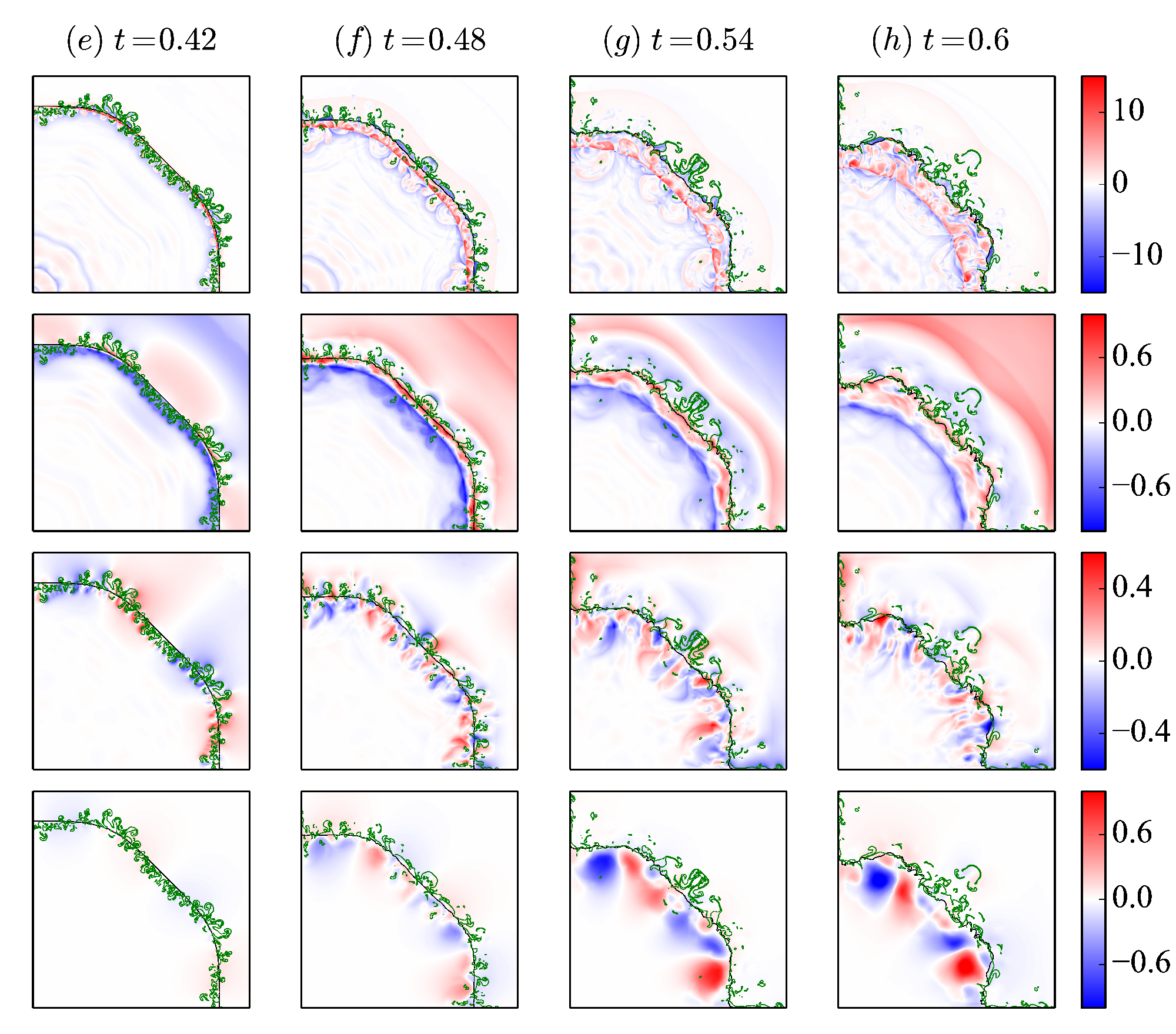}
\caption{\label{fig:fieldEvol_dd0_1} Evolution of charge density and electromagnetic field overlaid with ion interface (black line) and electron interface (green line). In each subplot, from top to bottom: charge ($\sum_\alpha n_\alpha q_\alpha$), radial electric field ($E_r$), azimuthal electric field ($E_\theta$) and magnetic field ($B_z$).}
\end{figure}
The nature of the two-fluid plasma model allows the self-consistent generation of electromagnetic fields even though these are zero initially. Since our simulations are confined to be two-dimensional simulations ($x-y$ plane), the electric field is generated in the $x-y$ plane while the self-generated magnetic field is along the $z$ axis. To illustrate the initial generation of the electromagnetic field, we focus on one time instant ($t=0.075$) when the first electron shock-interface interaction occurs. During the interaction, the azimuthal electron velocity induced by the vorticity deposited on the interface has opposite sign on both sides of the interface and decays away from the interface, as shown in Fig. \ref{fig:fieldGen_dd0_1}(e). On the other hand, the ions near the interface are almost unperturbed due to their large inertia. From the Fig. \ref{fig:fieldGen_dd0_1}(f), we see that the azimuthal velocity of ions is about three orders of magnitude smaller than that of electrons. Thus, according to the Eq. \ref{eq:electric}, the main contribution to the growth of azimuthal electric field $E_\theta$ stems from the electron part. Since $E_\theta$ near the interface is almost zero before impacted by the shock (not exactly zero due to the effect of light waves), the  azimuthal electric field generated at first shares the same distribution as the azimuthal electron velocity. Along the radial direction, the processes are becoming complex. Similar to the azimuthal case, the radial ion velocity is so weak that it contribute littles to the radial current density, and hence it is not considered here. When the first electron shock converges toward the electron interface, the inward moving electrons leads to the negative radial electric field $E_r$ according to the Eq. \ref{eq:electric}. Then, the induced negative $E_r$ exerts a positive acceleration on the electrons that slows the inward $u_{e,r}$ and even reverses the direction of $u_{e,r}$ in a relatively short duration:  in Fig. \ref{fig:fieldGen_dd0_1}(d), we note that the radial electron velocity varies from negative to positive along $r$ direction in the region behind the first electron shock (except behind the reflected shock, where electrons move outward due to the shock reflection). Finally, the positive $u_{e,r}$ leads to a positive current density that weakens the $E_r$ in turn and even changes its sign (to positive) given sufficient time. As a result, we see the similar but lagging distribution of $E_r$ compared to $u_{e,r}$ in the Fig. \ref{fig:fieldGen_dd0_1}(a). We should note that the electric force term dominates the Lorentz force at early time, thus we do not consider magnetic force contributions in the previous discussion. For instance, at $t=0.075$, the magnitude of $c E_r$ is about four orders larger than that of $\left|u_{e,\theta}B_z\right|$. Since the induced $E_\theta$ decays away from the interface while $E_r$ varies in azimuthal direction, the curl of $E$ is not zero and results in the generation of magnetic field in $z$ direction according to Eq. \ref{eq:magnetic}, as shown in Fig. \ref{fig:fieldGen_dd0_1}(c). 
\begin{figure}
\includegraphics[width=0.49\linewidth]{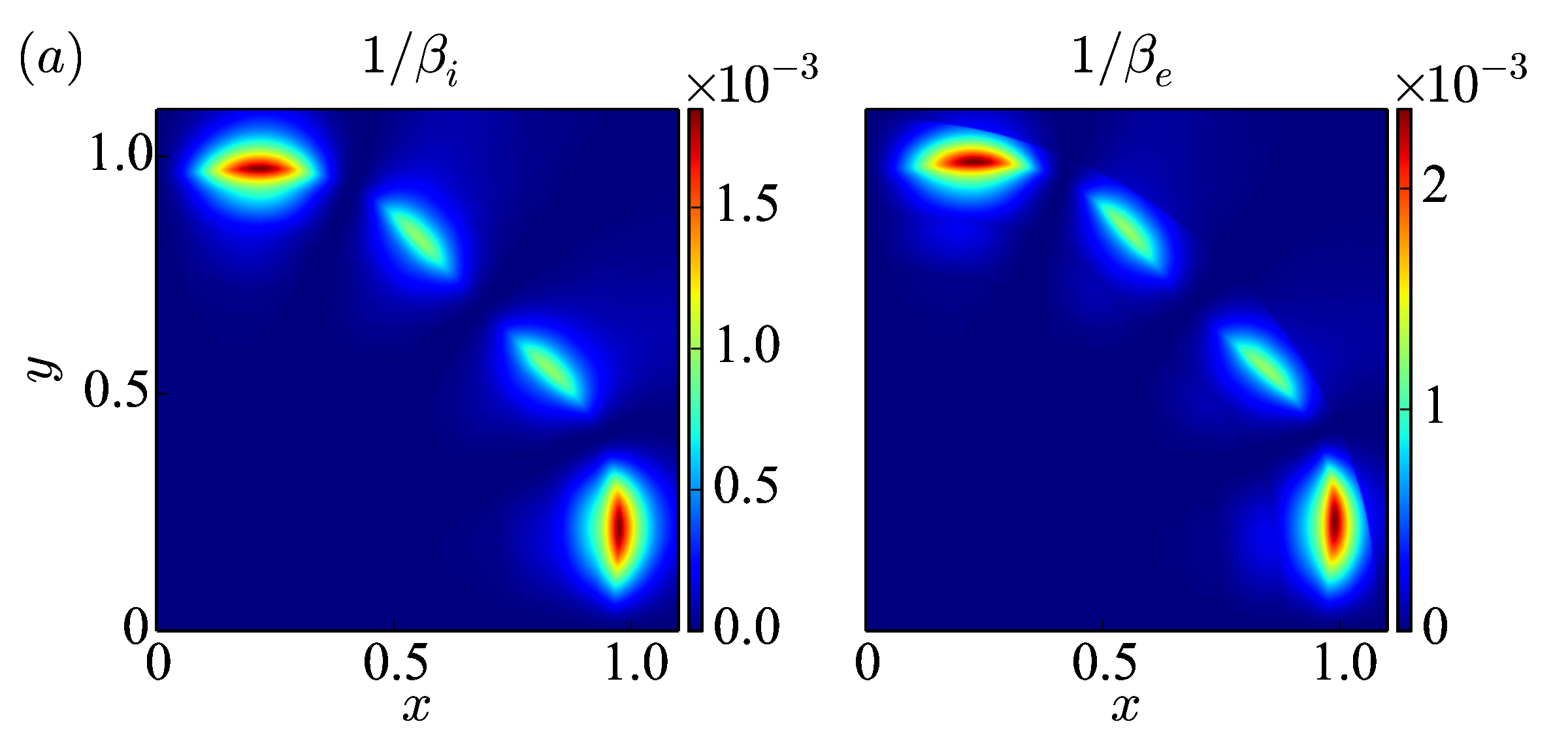}
\includegraphics[width=0.49\linewidth]{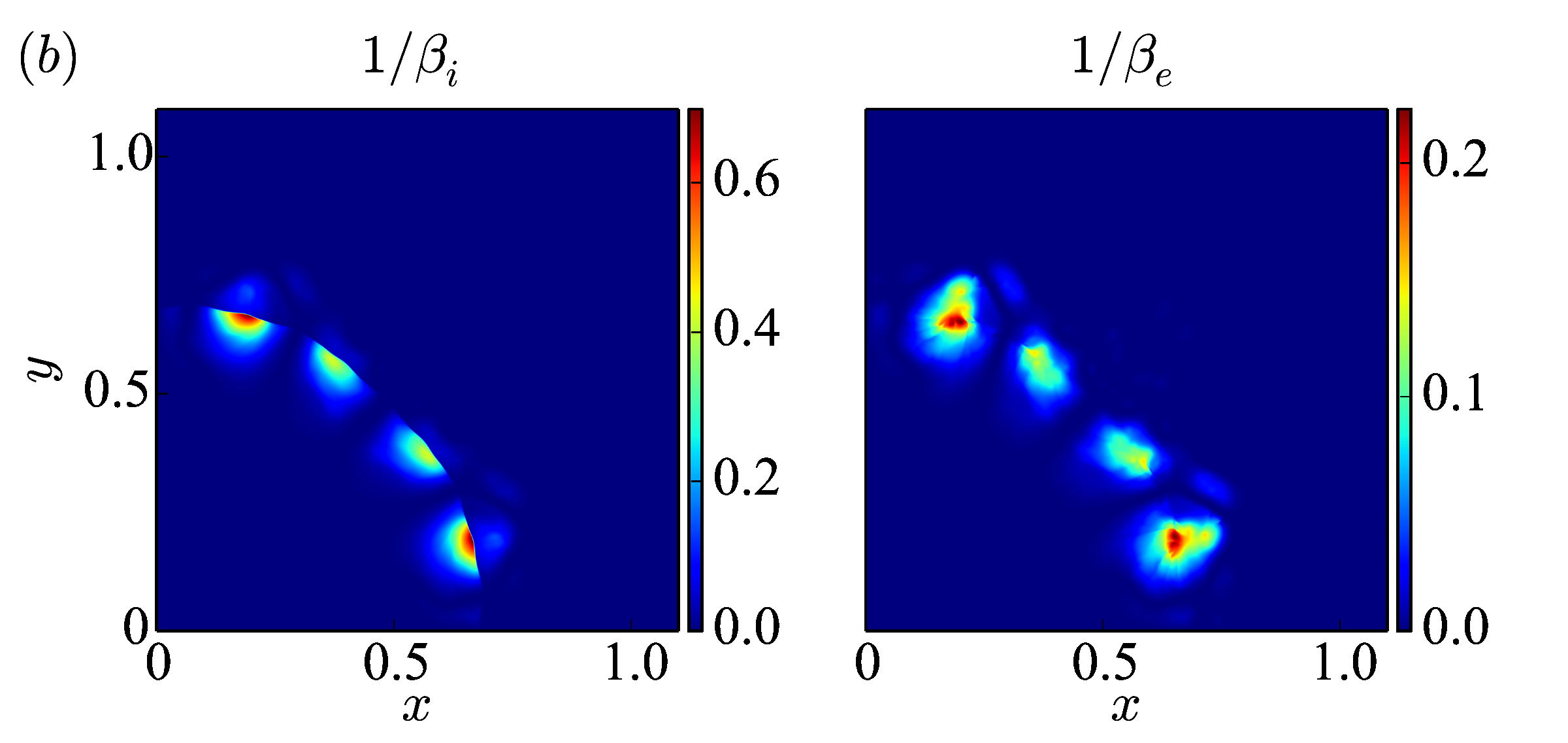}
\caption{\label{fig:invb_dd0_1} Local strength of the magnetic field, $1/\beta_\alpha=B_z^2/2p_\alpha$, at (a) $t=0.18$ and (b) $t=0.57$.}
\end{figure}
\begin{figure}
\includegraphics[width=0.49\linewidth]{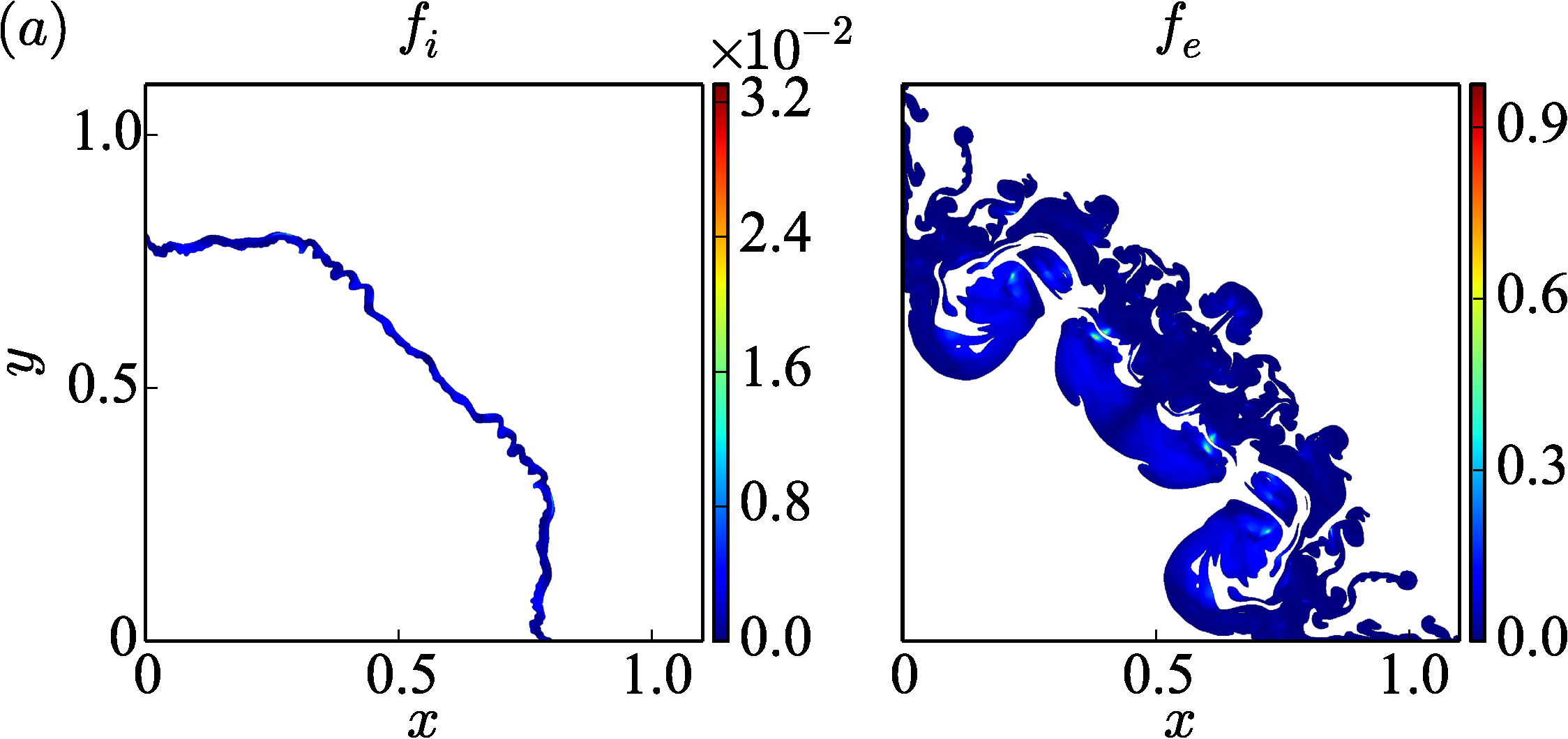}
\includegraphics[width=0.49\linewidth]{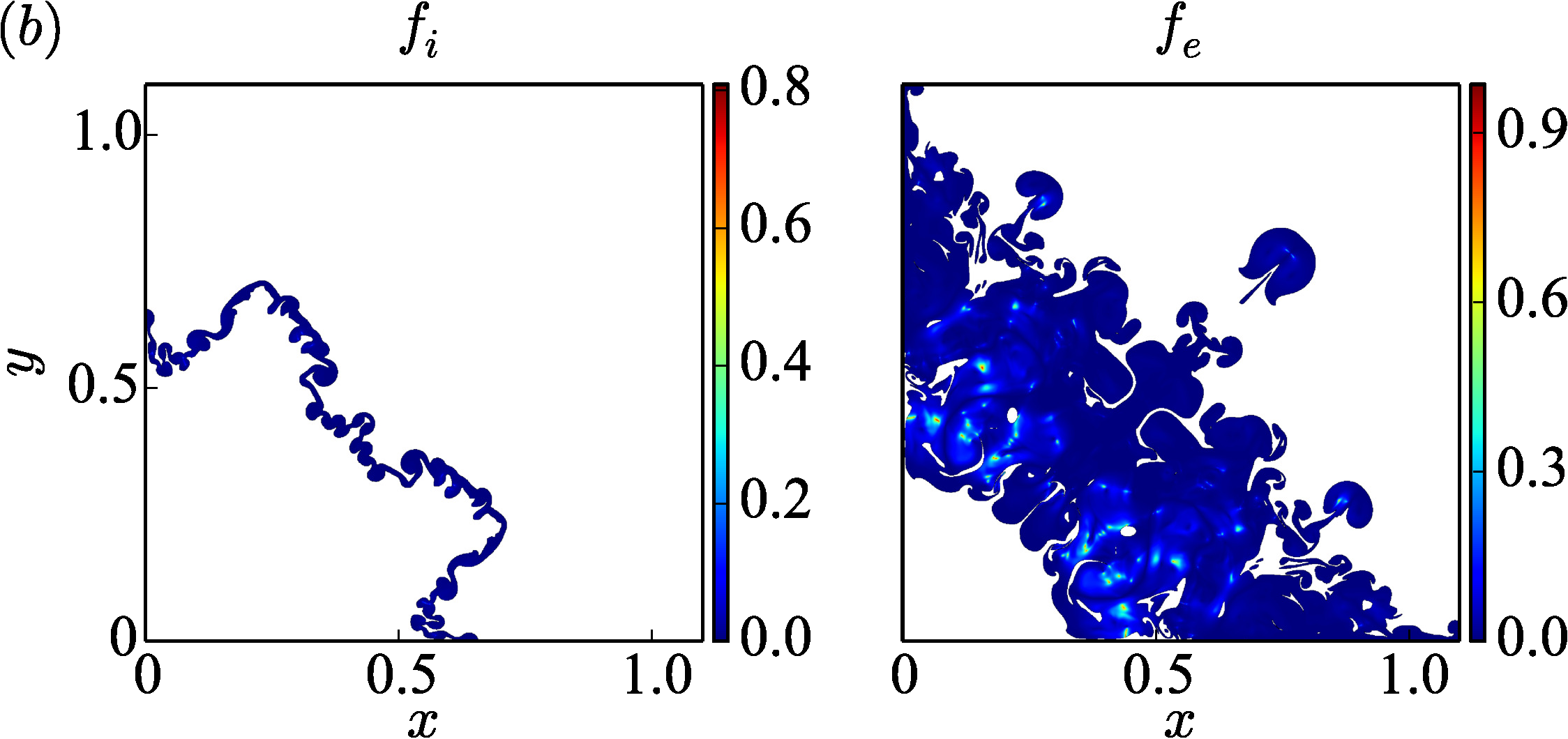}
\caption{\label{fig:magforce_dd0_1} Comparison between magnetic part ($\mathcal{L}_\alpha^M$) and electric part ($\mathcal{L}_\alpha^E$) of Lorentz force in the region of $-0.9 < \phi_\alpha < 0.9$ at (a) $t=0.57$ and (b) $t=0.75$, $f_\alpha=|\mathcal{L}_\alpha^M|/(|\mathcal{L}_\alpha^M|+|\mathcal{L}_\alpha^E|)$. }
\end{figure}
The evolution of electromagnetic field and charge density is plotted in Fig. \ref{fig:fieldEvol_dd0_1}. At $t=0.09$, the first electron shock has traversed the interface while the second one has not arrived yet. The imploding shocks generate a high electron density region behind the shocks, exhibiting  a band of negative charge density followed by an area of positive charge density for charge balance. During the time  interval $t=0.075-0.09$, the electrons near the interface move outward under the effect of outward electric force, resulting in the electron interface overshoot to the right of the ion interface, as shown in Fig.\ref{fig:DI_dd0_1}(b). In turn, the outward radial motion of the electrons results in a sign change of the radial electric field $E_r$ from negative to positive, as shown in Fig. \ref{fig:fieldEvol_dd0_1}(a). Before the ion shock arrives, multiple electron shock-interactions occur and form bands with alternate positive and negative variations in charge and radial electric field. The oscillations of $E_r$ on the interface causes oscillations of the $r$-direction acceleration of the electrons that results in an the oscillatory movement of the electron interface, as previously discussed in \ref{sec:flowevol}. The oscillation of electric field at the interface, results in a magnetic field on the interface that varies over time regularly. During the ion shock-interface interaction, the secondary instabilities on the electron interface rapidly grow due to a long-lived electron acceleration along RT unstable direction (compared with the oscillating electron acceleration produced by the electron shocks) and distorts the electron interface. As a consequence, the area of charge separation becomes chaotic and  spans on either side of the ion shock at $t=0.48$. The same physical effects lead to a chaotic distribution of electromagnetic field that spreads over the region of charge separation. The local strength of the self-generated magnetic field, $1/\beta_\alpha = B_z^2/2p_\alpha$ is plotted in Fig. \ref{fig:invb_dd0_1}. At $t=0.18$, although multiple electron shock-interface interactions have occurred, the generated magnetic field is still quite weak for both two fluids with the minimum $\beta=541 (424)$  for ions (electrons). Thus, the magnetic field has a somewhat insignificant effect on the evolution of the flow field during the time. After the ion shock impacted on the interface, the magnetic field grows sufficiently large that its influence on the flow evolution cannot be ignored. For instance, at $t=0.57$, the minimum $\beta\approx 4.5(1.4)$ for electrons (ions). We note that the generated magnetic field is not aligned in the appropriate direction that can suppress the RM instability as in the MHD \citep{wheatley2005, wheatley2014}. But it does plays a vital role in the electromagnetically driven RT instability at late time. Fig. \ref{fig:magforce_dd0_1} compares the magnetic part and electric part of Lorentz force in the region where the tracer $\phi_\alpha \in(-0.9,0.9)$. We use $f_\alpha=|\mathcal{L}_\alpha^M|/(|\mathcal{L}_\alpha^M|+|\mathcal{L}_\alpha^E|)$ to quantify the contribution of magnetic force. The magnetic force becomes trivial as $f_\alpha$  tends to $0$, and vice versa. At $t=0.57$, the magnetic force is negligible in the ions since the maximum $f_i \approx 0.033$. While in the electrons, the maximum $f_e \approx 0.98$ and the magnetic force is comparable to the electric part in a considerable area. Thus, the magnetic field has an influence on electromagnetically driven RT instability in electrons. At $t=0.75$, the maximum $f_i$ increases to $\approx 0.81$ implying that the magnetic force can't be ignored in ions. Therefore, in the ions, although the generated magnetic field has little direct effect on the electromagnetically driven RT instability at $t=0.57$, it may play an important role at $t=0.75$.
\subsection{\label{sec:smallDebye}Reduced length-scale case}
\begin{figure}
\includegraphics[width=1\linewidth]{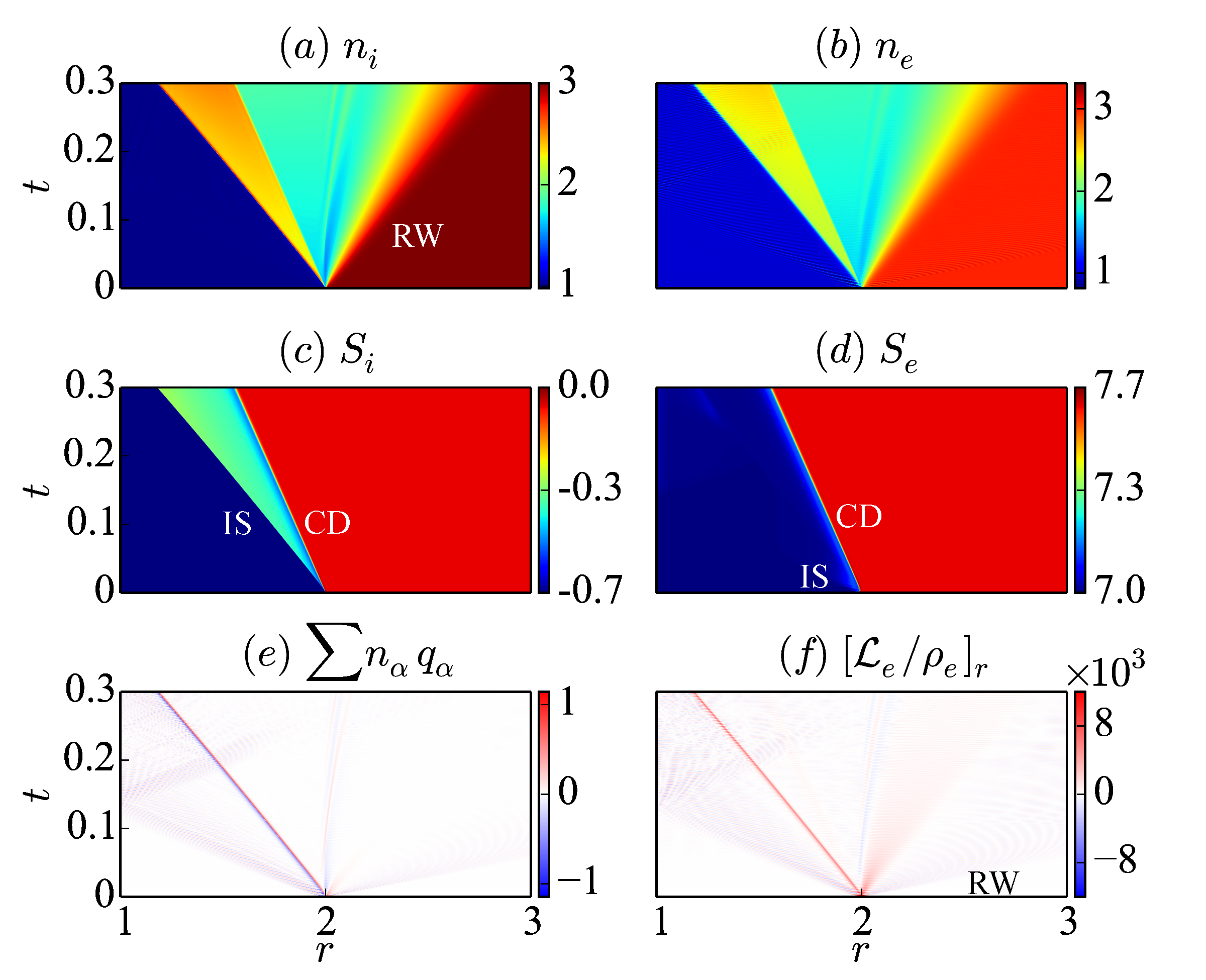}
\caption{\label{fig:xt_dd0_01} Space-time diagrams along $r$ direction with $d_{D,0}=0.01$. (a) ion number density, (b) electron number density, (c) ion entropy $S_i=\ln{(p_i/\rho_i^\gamma)}$, (d) electron entropy $S_i=\ln{(p_e/\rho_e^\gamma)}$, (e) charge density, (f) Lorentz acceleration of electrons in $r$.}
\end{figure}
\begin{figure}
\includegraphics[width=1\linewidth]{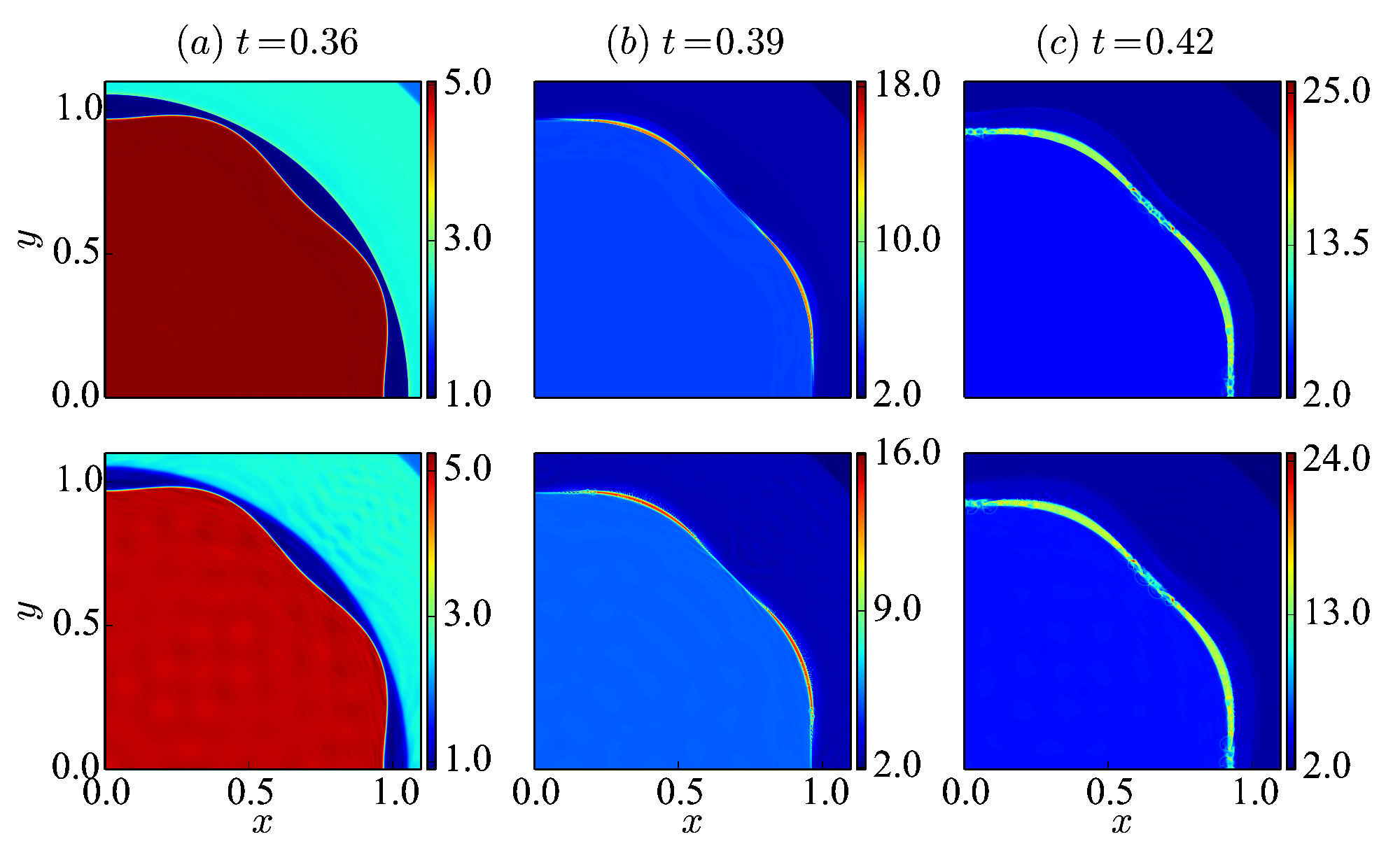}\\
\includegraphics[width=1\linewidth]{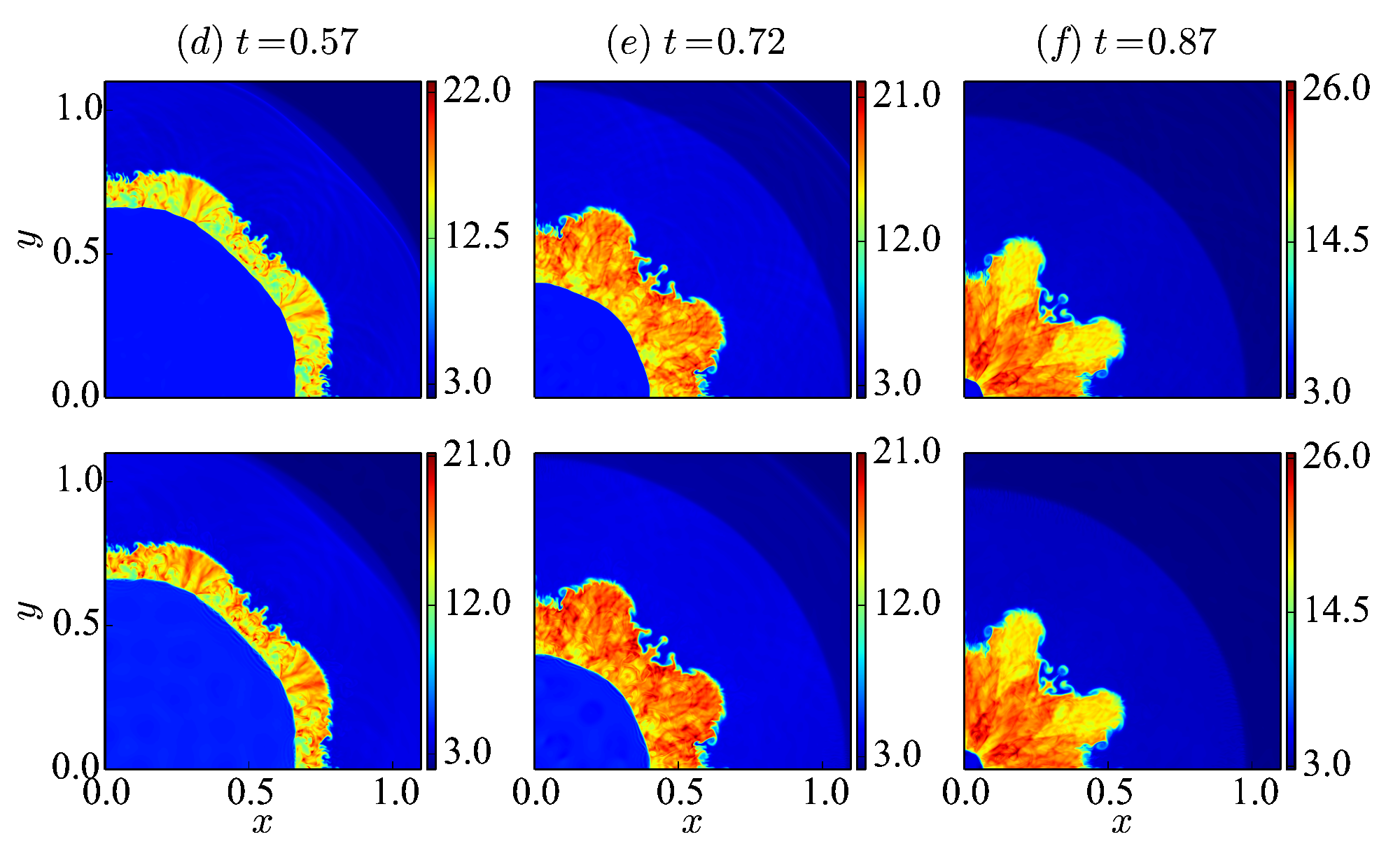}
\caption{\label{fig:numden_dd0_01} Number density of ions (top) and electrons (bottom) at various time with $d_D=0.01$. (a) $t=0.36$, (b) $t=0.39$, (c) $t=0.42$, (d) $t=0.57$, (e) $t=0.72$, (f) $t=0.87$}
\end{figure}
\begin{figure}
\includegraphics[width=\linewidth]{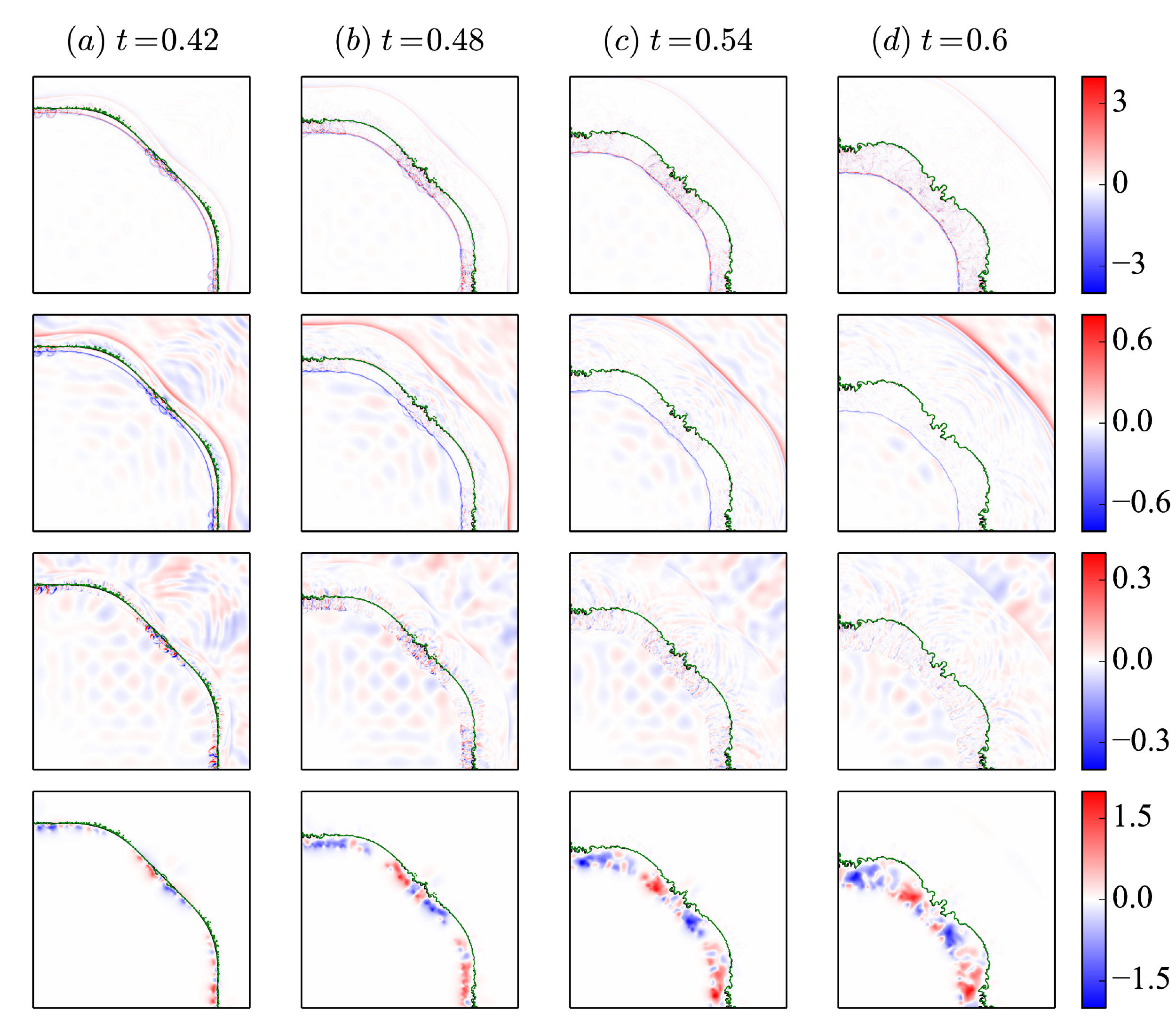}
\caption{\label{fig:fieldEvol_dd0_01} Evolution of charge density and electromagnetic field overlaid with ion interface (black line) and electron interface (green line). In each subplot, from top to bottom: charge ($\sum_\alpha n_\alpha q_\alpha$), radial electric field ($E_r$), azimuthal electric field ($E_\theta$) and magnetic field ($B_z$).}
\end{figure}
The ideal MHD limit of two-fluid plasma model is achieved by taking the following limits: light speed $c \rightarrow \infty$ and $d_{D,0}c \rightarrow 0$. Presently, we keep the light speed fixed, and approach the ideal MHD model by decreasing one order magnitude of the reference Debye length $d_{D,0}$. Through the simulations with $d_{D,0}=0.01$, we study the flow evolution and instabilities growth when the two-fluid plasma tends to ideal MHD. According to the Eq. \ref{eq:momentum}, the Lorentz force is inversely proportional to $d_{D,0}$ ($\mathcal{L}_\alpha \sim 1/d_{D,0}$). Thus, the coupling between ions and electrons becomes stronger as the $d_{D,0}$ decreases. To illustrate the wave structures generated from the Riemann interface in this case, we plot the space-time diagram for $t\le 0.3$ in Fig. \ref{fig:xt_dd0_01}. At first glance, the electron wave structures are much cleaner compared to the previous case, with no discernible oscillations in $n_e$ here compared with those for higher $d_{D,0}=0.1$ (Fig. ~\ref{fig:xt_dd0_1}). The structures of number density between ions and electrons are similar since the electrons are tightly coupled to the high-inertia ions. As a result, the extent of charge separation region is small and localized to the shocks and rarefaction waves. However, the magnitudes of charge separation are comparable between the two cases, as shown in Fig. \ref{fig:xt_dd0_01}(e). In addition, the electron Lorentz acceleration becomes one order magnitude larger due to the one order magnitude larger Lorentz force induced here, which implies that the generated electromagnetic field are of the same order of magnitude in the two cases. The extent of the Lorentz force acting area is much smaller, consistent with the localized charge separation region. In fact, the multiple electron shocks generated from the Riemann interface still exist in this case, however the shock strengths decrease over a small distance and degenerate to the highly oscillatory wave packets, due to the larger restoring Lorentz force acting on them during the convergence. Thus, unlike the large $d_{D,0}$ case (Fig. \ref{fig:xt_dd0_1}(d)), here we do not observe the large scale layers caused by the electron shocks, see Fig. \ref{fig:xt_dd0_01}(d). It is noted that the oscillations of the contact discontinuity in electrons are observed in this case but with smaller amplitudes and higher frequency. These oscillations are rapidly damped due to the larger Lorentz force. In general, the coupling effect between ions and electrons decreases the converging speed of electron waves while increases that of ion waves. Thus, the electron waves move slower while ion shock moves faster for the smaller $d_{D,0}$ case. We see that the first electron wave reaches interface at a later time than that of the previous case, while we note opposite effect for the ions. For instance, the ion shock arrives at the interface at $t \approx 0.36$ in this circumstance, compared to $0.39$ for the case with $d_{D,0}=0.1$. Fig. \ref{fig:numden_dd0_01} shows the evolution of number density of ions and electrons for $d_{D,0}=0.01$ case while the charge density and induced electromagnetic field are shown in Fig. \ref{fig:fieldEvol_dd0_01}. Obviously, the difference in density field morphology between ions and electrons is much smaller at each time compared to the higher $d_{D,0}=0.1$ case as in  Fig. \ref{fig:numden_dd0_1}, resulting in the localized charge separation zone. Before the ion shock-interface interaction occurs, multiple electron waves have travelled across the electron interface at $t=0.36$, the electrons are disturbed but not significantly, since the incoming electron waves are weak. After the ion shock-interface interaction, the transmitted ion shock is highly perturbed by the interface and electromagnetic force, resulting in the formation of kinks and shock-shock interactions as seen in the chaotic region behind the shock. During the time, both RM and RT instabilities  are manifested for the ion interface. In general, the magnitude of Lorentz acceleration acting on the ion interface is larger than that in the previous case, however, the region of the acceleration on the ion interface decreases due to the localized electromagnetic fields and the direction of the acceleration significantly varies around the interfaces due to the chaotic flow, as a result, the electromagnetically driven RT instability is reduced leading to a smaller extent of the perturbation growth. In addition, the growth of the instabilities on the electron interface is restrained since the electrons are strongly coupled to the ions. 
\begin{figure}
\includegraphics[width=1\linewidth]{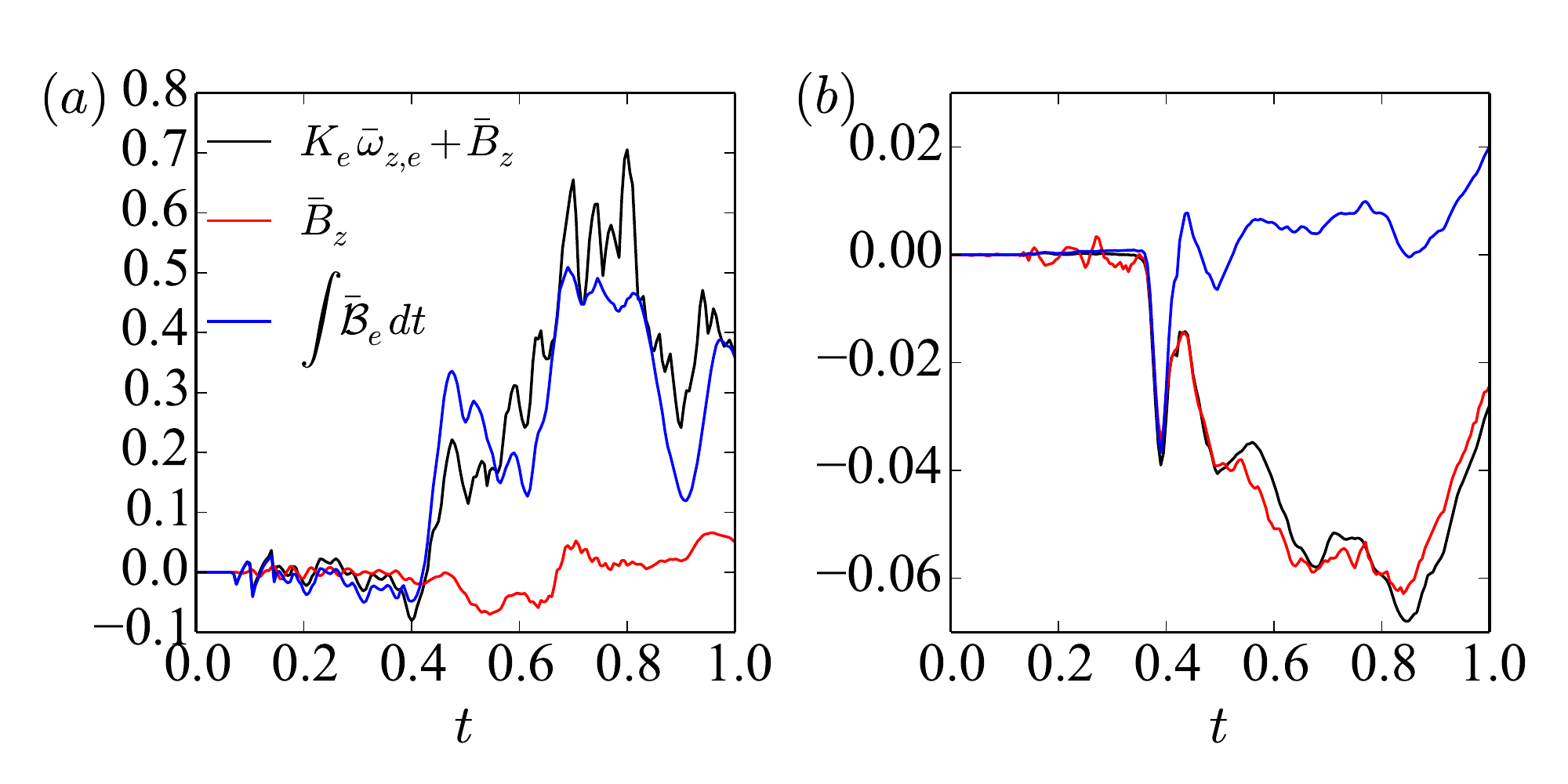}
\caption{\label{fig:biermann}  Evolution of self-generated magnetic field $\bar{B}_z$, $K_\alpha \bar{\omega}_{z,e}+\bar{B}_z$ and the cumulative time integral of Biermann battery term $\bar{\mathcal{B}}_e$. All quantities are averaged over the region $S$: $\left(r,\theta\right)\in \left(0,1.1\right)\times \left(3\pi/4, \pi/2\right)$. (a) $d_{D,0}=0.1$, (b) $d_{D,0}=0.01$.}
\end{figure}

Applying the curl operator to the momentum equation Eq. \ref{eq:momentum}, and substituting it into the induction equation Eq. \ref{eq:magnetic} (ignore the divergence mitigation term, $\Gamma_B=0$), we get an alternative form of magnetic field equation,
\begin{equation}
\frac{\partial}{\partial t}\left(\bm{B}+K_e\bm{\omega}_e\right)=K_e\frac{\nabla \rho_e \times \nabla p_e}{\rho_e^2} +\nabla \times \left(\bm{u}_e \times \left(\bm{B}+K_e\omega_e\right)\right),
\label{eq:magnetic_new}
\end{equation}
where $K_e=c d_{D,0} m_e/q_e$ and $\bm{\omega}_e=\nabla \times \bm{u}_e$ is the electron vorticity. The first term on the right hand side, proportional to the baroclinic contribution and referred to as the Biermann battery effect \citep{biermann1950}, is considered to be the source of the self-generated magnetic field, denoted as $\mathcal{B}_e$ here. The second term exhibits the combined effects of the convection and stretching or tilting of the vorticity and magnetic field. The equation shares the same form as the HD vorticity equation except that the vorticity is replaced by the generalized vorticity $\bm{\omega}_e+\bm{B}/K_e$ here. In the ideal MHD limit, $K_e$ tends to $0$, the equation degenerates to the Eq. \ref{eq:magnetic_mhd} with no self-generated magnetic field allowed. 
\begin{equation}
\frac{\partial \bm{B}}{\partial t}=\nabla \times \left( \bm{u} \times \bm{B}\right).
\label{eq:magnetic_mhd}
\end{equation}
To study the influence of the Biermann battery term on the net magnetic field generation, we average $K_e \omega_{z,e} + B_z$, $B_z$ and $\int \mathcal{B}_e\,dt$ over the region $S$: $\left(r,\theta\right)\in \left(0,1.1\right)\times \left(3\pi/4, \pi/2\right)$. The results are plotted in the Fig. \ref{fig:biermann}. For $d_{D,0}=0.1$ case, $\bar{B}_z$ is not the major contributor to $K_e\bar{\omega}_{z,e}+\bar{B}_z$. During the first converging electron shock-interface interaction that begins at $t\approx 0.07$, $\nabla \rho_e \times \nabla p_e$ becomes positive on the electron interface in region $S$, leading to the negative $\mathcal{B}_e$ at the time. As a consequence, $\int\bar{\mathcal{B}}_e dt$ decreases to $-0.02$ during the the very short period, see in Fig. \ref{fig:biermann}(a). Then, as discussed in \ref{sec:flowevol}, the flow field exhibits spatio-temporal oscillations  under the combined effects of the multiple electron shock-interface interactions and the induced Lorentz force, resulting in the oscillation of $\nabla \rho_e \times \nabla p_e$, same as the $\int\bar{\mathcal{B}}_e\, dt$. During the ion shock-interface interaction period that begins at $t\approx 0.39$, the $\nabla \rho_i \times \nabla p_i$ is positive on the ion interface, causing the positive of $\nabla \rho_e \times \nabla p_e$ at the corresponding region due to the coupling effect. Hence, we can see $\int\bar{\mathcal{B}}_e\, dt$ decreases to $-0.048$ at $t\approx 0.4$. Soon, the negative $\nabla \rho_e \times \nabla p_e$ begins to take control in region $S$ and $\int\bar{\mathcal{B}}_e\, dt$ rapidly grows into and keeps positive for the rest of the simulation time. It is noted that the following complex behavior of $\int\bar{\mathcal{B}}_e\, dt$ comes from the complicated structures of the flow field in $S$. We see that the Biermann battery term is the primary contribution to $K_e\bar{\omega}_{z,e}+\bar{B}_z$ during the simulation time. However, the effects from the second term cannot be ignored due to the high mobility of the electrons, especially after the ion shock-interface interaction, when the electrons are highly convected and chaotic. Therefore, both Biermann battery term and the second term effect dominate the net magnetic field in $S$. For $d_{D,0}=0.01$ case, the magnetic field part dominates $K_e\bar{\omega}_{z,e}+\bar{B}_z$ during the simulation time, in contrast to the previous case. We see that the cumulative Biermann battery effect is almost zero before $t\approx 0.36$, since the converging electron waves are so weak that the induced baroclinic term is negligible. In addition, during the period, the nearly zero $K_e\bar{\omega}_{z,e}+\bar{B}_z$ indicates little contribution from the second term. However, it allows oppositely signed vorticity part and magnetic field part that may cancel each other out. Therefore, other than the Biermann battery effect, the second term is the source of the net magnetic field and vorticity in $S$. After the ion shock-interface interaction, the coupling effect induces the positive baroclinic term in electrons, leading to the decreasing of $\int\bar{\mathcal{B}}_e\, dt$ to $\approx -0.037$ at $t=0.39$, then the negative baroclinic term dominates and  $\int\bar{\mathcal{B}}_e\, dt$ increases. During the period from $0.37$ to $0.4$, the apparently enlarged $\int\bar{\mathcal{B}}_e\, dt$ is the main contribution to $\bar{B}_z$, after that, the effects from second term dominate again for the net magnetic field. 
\subsection{\label{sec:perturbations}Interface stastitics}
\begin{figure}
\includegraphics[width=1\linewidth]{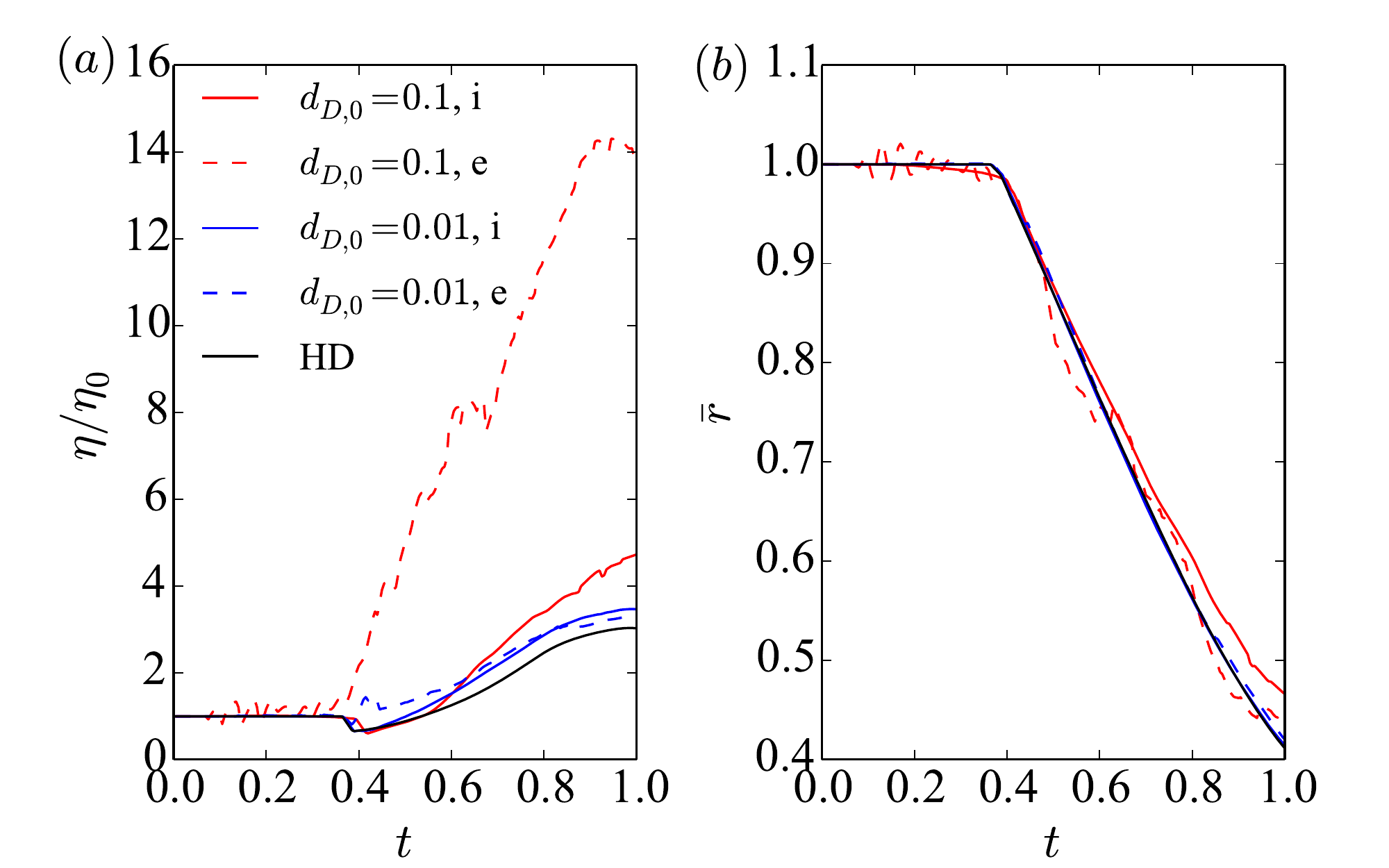}
\caption{\label{fig:ampEvol}  Evolution of (a) relative perturbation amplitudes, $\eta/\eta_{0}$ and (b) averaged interface positions, $\bar{r}$. `i' denotes ions, `e' denotes electrons, `HD' denotes the corresponding hydro-case.}
\end{figure}
\begin{figure}
\includegraphics[width=1\linewidth]{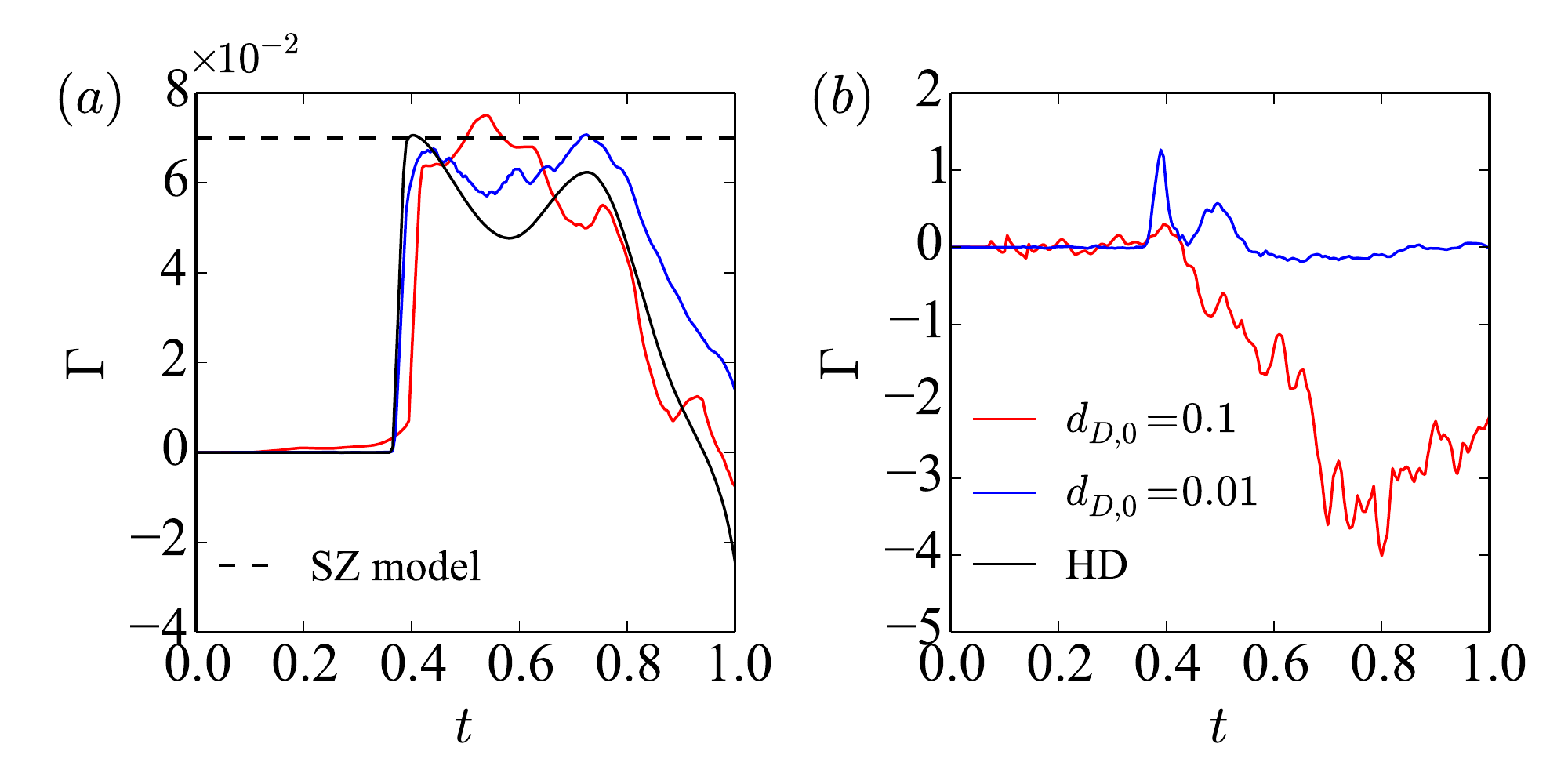}
\caption{\label{fig:vorEvol}  Evolution of (a) ion and (b) electron circulations with various $d_{D,0}$. `HD' is the corresponding hydro-case. `- -' is the predicted circulation of HD case with SZ model.}
\end{figure}
To investigate the evolution of perturbation amplitude, we calculate the mean volume-of-fluid tracer $\Phi_\alpha$ along the azimuthal direction,
\begin{equation}
\Phi_\alpha\left(r\right)=\int_0^{\pi/2} \phi_\alpha\left(r,\theta\right) d\theta.
\label{eq:Phi}
\end{equation}
The perturbation amplitude $\eta_\alpha$ is defined as the width of the region where $0.05 < \Phi_\alpha < 0.95$. Fig. \ref{fig:ampEvol}(a) shows relative perturbation amplitudes of the two fluids, normalized by the initial amplitude $\eta_0$. We also consider the motion of the interfaces, as shown in Fig. \ref{fig:ampEvol}(b). The mean position of the interface $\bar{r}_\alpha$ is defined at the center of the region where $0.05 < \Phi_\alpha < 0.95$. The results are compared with those of the corresponding HD case. It shows that the instability of the interface is amplified in the two-fluid case. For $d_{D,0}=0.1$ case, since the coupling effect is not strong, the evolution of the perturbation amplitude is significantly different between the two fluids. The peaks of the relatively perturbation amplitudes of ions and electrons are $4.72$ and $13.78$, respectively, in comparison with $3.03$ of the HD case. The amplified ion perturbation amplitude is attributed to the electromagnetically driven RT instability, as discussed in \ref{sec:flowevol}. Fig. \ref{fig:ampEvol}(b) shows that the electron interface oscillates around the ion interface under the effects of the multiple electron shocks and Lorentz force for $t\le 0.39$, in accordance with the discussions in \ref{sec:flowevol}. Though the electron interface oscillates, it slowly converges. As a consequence, the ion interface slowly converges due to the coupling effect. During this time, the electron perturbations are influenced by the compression effect from shocks, RT effect and RM unstable effect, leading to the oscillation of amplitude, as shown in Fig. \ref{fig:ampEvol}(a), while the ion perturbation amplitude remains almost  unchanged. After the ion shock-interface interaction occurs, the perturbations start to grow and develop into RM and RT instabilities. For the ion part, the amplitude increases and attains its maximum growth rate at $t\approx 0.56$. The acceleration and deceleration of the ion interface appear alternately and lead to the alternate RT unstable and stable effects on the ion perturbations, respectively. As a consequence, the ion perturbation amplitude growth is not as temporally smooth as the HD case. Compared with the ions, the electron perturbations grow more rapidly due to their lighter mass, and the evolution of interface mean position is complicated because of the significant influence exerted by the Lorentz force. As the reference Debye length is reduced, the coupling effect becomes strong so that the electrons are constrained to the ions. Therefore, we see the interfaces in both fluids move together without distinct divergence. Since this case is closer to the single-fluid MHD limit, we see that the evolution of ion interface mean position matches well with the HD case, in contrast to the previous higher $d_{D,0}=0.1$ case. The ion perturbation amplitude coincides with HD case for early times, then exceeds it due to the electromagnetically driven RT instability, however, the maximum value is $3.47$, much closer to the HD case than the previous case. Since the electron converging waves are too weak to influence the electron perturbations, the perturbation amplitude remains unchanged before ion shock arrives. Though there is an obvious amplification of the electron perturbation amplitude after ion shock-interface interaction, the electrons are so restrained that the maximum amplitude is $3.29$, even less than for the ion interface. 

The vorticity plays an important role in perturbation growth. In this paper, we consider the region where $0.025<\phi_\alpha<0.975$ as the inner part of the interface. Then, the ion and electron circulation on a half period of the perturbed interface is defined as,
\begin{equation}
\Gamma_\alpha=\int_{3\pi/4}^{\pi/2}\int_{0.025}^{0.975} \omega_{z,\alpha} d\phi_\alpha d\theta.
\label{eq:circulation}
\end{equation}
Fig. \ref{fig:vorEvol} plots the evolution of the ion and electron circulation for the cases with $d_{D,0}=0.1$ and $0.01$. The ion circulations are compared with the result of HD case. In HD case, the approximate Mach number of the incident shock wave arriving at the interface is 2. With the SZ model provided by the Samtaney \& Zabusky \citep{samtaney1994}, the total circulation deposited on the interface is predicted to be $7.0 \times 10^{-2}$, which agrees well with the simulation result. As $d_{D,0}$ decreases, the ion circulation evolution tends to the HD case. For $d_{D,0}=0.1$ case, the oscillation of the electron circulation $\Gamma_e$ at early time comes from the competitive mechanism between the deposited baroclinic vorticity and Lorentz induced vorticity, in accordance with the oscillation of $\eta_e$. Later, the vorticity generated by the RT effect of the interface motion dominates and $\Gamma_e$ decreases to be negative, while the vorticity induced by the Lorentz force causes disturbances to the curve. We see the ion circulation accumulates before the ion shock arrives, however, it is too small to amplify $\eta_i$, as shown in Fig. \ref{fig:ampEvol}. During the ion shock passage through the interface, the baroclinic vorticity depositing on the interface leads to the rapid increase of ion circulation. Unlike the HD case, the subsequent increase of $\Gamma_i$ occurs due to the acceleration of the ion interface. As a consequence, $\Gamma_i$ achieve its maximum at $t\approx0.54$, resulting in the maximum growth rate of $\eta_i$ around the time. After that, though the acceleration of the ion interface increases $\Gamma_i$ at $t\approx0.73$ and $0.89$ for a short period, overall, the deceleration of the interface and the convection of the vorticity dominates and reduces $\Gamma_i$. For $d_{D,0}=0.01$ case, before ion shock arrives, both electron and ion circulation are almost $0$, leading to the almost unchanged amplitudes in Fig. \ref{fig:ampEvol}. After that, a significant growth of electron circulation $\Gamma_e$ appears due to the induced Lorentz force, resulting in the obvious amplification of $\eta_e$ as shown before. The mismatch of the perturbation growth between ions and electrons overturns the Lorentz force, causing the rapid decrease of $\Gamma_e$. Following that, the acceleration of electron interface increases $\Gamma_e$ again, as well as the distinction between ion and electron interfaces until  the vorticity generated by Lorentz force dominates and reduces $\Gamma_e$ again. Such a competing mechanism occurs all the time, as a consequence the growth of electron circulation and the perturbations being constrained. For the ion fluid, the $\Gamma_i$ curve tends to HD case, however the influences of ion interface motion and Lorentz force are still significant.
\section{\label{sec:conclusions}Conclusion}
In summary, the evolution of interfacial instabilities in a cylindrically imploding flow is examined in the framework of an ideal two-fluid plasma model. A Riemann interface is used to generate the converging shocks with one shock in ions and multiple shocks in electrons. Due to the smaller electron mass, electron shocks travel faster than the ion shock. When the Debye length is large, i.e., the coupling effect is small, the flow fields differ significantly between ions and electrons. Before the ion shock-interface interaction occurs, the electron density interface oscillates under the compression effect of the electron shocks and the restoring effect of the incident Lorentz force while the ion interface is nearly still. During that time, the spatio-temporal varying Lorentz force drives the secondary RT instability on the electron interface and it develops fine-scale structures. In addition, the electron vorticity on the interface oscillates with a small amplitude, resulting in the oscillation of the perturbation amplitude. In contrast to the electron shocks, the ion shock-interface interaction not only tremendously increases the vorticity magnitude that leads to the RM instability, but also creates a long-lived Lorentz acceleration on the interfaces in the RT unstable direction, resulting in the RT instability. The combination of RM and electromagnetic driven RT instabilities lead to larger perturbation amplitudes than that of HD case. Moreover, the Biermann battery effect is the main source of self-generated magnetic field for the entire duration. As the Debye length decreases, the coupling between the electron and ion fluids is stronger and the electrons are tightly bound to the ions. Therefore, the flow fields between ions and electrons tend to be the same and be closer to the HD case. In this circumstance, the electron shocks from the Riemann interface are weakened quickly due to the stronger restoring Lorentz force and degenerate into wave packets with a resulting weaker effect on the electron interface. The more localized and perturbed self-generated electromagnetic field on the interface leads to lesser extent of electromagnetic driven RT instability. As a consequence, the perturbation amplitude reduces compared with the large Debye length case, though it is still larger than that of HD case. Unlike the larger Debye length case, the Biermann battery effect contributes mainly to the magnetic field during the ion shock-interface interaction. 
\section{Acknowledgements}
This research was supported by the KAUST Office of Sponsored Research under Award URF/1/3418-01.
\nocite{*}
\bibliography{references}

\end{document}